\documentclass[bibyear]{aa}
\usepackage{graphicx}
\usepackage{txfonts}
\usepackage{natbib,twoopt}
\usepackage[breaklinks=true]{hyperref} 
\bibpunct{(}{)}{;}{a}{}{,}             
\makeatletter
  \newcommandtwoopt{\citeads}[3][][]{\href{http://adsabs.harvard.edu/abs/#3}%
    {\def\hyper@linkstart##1##2{}%
     \let\hyper@linkend\@empty\citealp[#1][#2]{#3}}}
  \newcommandtwoopt{\citepads}[3][][]{\href{http://adsabs.harvard.edu/abs/#3}%
    {\def\hyper@linkstart##1##2{}%
     \let\hyper@linkend\@empty\citep[#1][#2]{#3}}}
  \newcommandtwoopt{\citetads}[3][][]{\href{http://adsabs.harvard.edu/abs/#3}%
    {\def\hyper@linkstart##1##2{}%
     \let\hyper@linkend\@empty\citet[#1][#2]{#3}}}
  \newcommandtwoopt{\citeyearads}[3][][]%
    {\href{http://adsabs.harvard.edu/abs/#3}
    {\def\hyper@linkstart##1##2{}%
     \let\hyper@linkend\@empty\citeyear[#1][#2]{#3}}}
\makeatother

\begin{document}

   \title{Emission from HCN and CH$_3$OH in comets}

   \subtitle{Onsala 20-m observations and radiative transfer modelling}

   \author{
          P. Bergman\inst{1}
          \and
          M. S. Lerner\inst{1}
          \and
          A. O. H. Olofsson\inst{1}
          \and
          E. Wirstr{\"o}m\inst{1}
          \and
          J. H. Black\inst{1}
          \and
          P. Bjerkeli\inst{1}
          \and
          R. Parra\inst{2}
          \and
          K. Torstensson\inst{2}
          }
        \institute{
          Department of Space, Earth and Environment, Chalmers University of Technology, Onsala Space Observatory, 43992 Onsala, Sweden\\
          \email{per.bergman@chalmers.se}
         \and
          European Southern Observatory, Av. Alonso de Cordova 3107, Vitacura, Santiago, Chile\\
        }

   \date{Received ; accepted }
 
  \abstract
   {}
   {The aim of this work is to characterize HCN and CH$_3$OH emission from recent comets.
     }
   {We used the Onsala 20-m telescope to search for millimetre transitions of HCN towards a sample of 11 recent and mostly bright comets
   in the period December 2016 to November 2019. Also CH$_3$OH was searched for in two comets. The HCN sample includes the
   interstellar comet 2I/Borisov. For the short-period comet 46P/Wirtanen we could monitor the variation of HCN emission over a time span of about one month.
    We performed radiative transfer modelling for the observed molecular emission by also including time-dependent effects due to the outgassing of molecules.
    }
   {HCN was detected in 6 comets. Two of these are short-period comets and 4 of them are long-period. Six methanol transitions were
   detected in 46P/Wirtanen, enabling us to determine the gas kinetic temperature. From the observations, we determined the molecular production rates using time-dependent radiative transfer modelling. For 5 comets, we could determine that the
   HCN mixing ratios lie near 0.1\% using contemporary water production rates, $Q_\mathrm{H_2O}$, taken from other studies.
   This HCN mixing ratio
   was also found typical in our monitoring observations of 46P/Wirtanen but here we notice deviations, on a daily time scale, up to 0.2\% which could indicate short-time changes in the outgassing activity.
   From our radiative transfer modelling of cometary comae, we found that time-dependent effects on the HCN level populations are of the order 5-15\%
   when $Q_\mathrm{H_2O}$ is around $2\times 10^{28}\,\mathrm{mol\, s^{-1}}$. The effects may be relatively stronger for comets
   with lower $Q_\mathrm{H_2O}$. The exact details of the time-dependent effects depend on the amount of neutral and electron collisions,
   radiative pumping, and molecular parameters such as the spontaneous rate coefficient.
   }
   {}
   
               
               \titlerunning{Emission from HCN and CH$_3$OH in comets}
               \authorrunning{Bergman et al.}

   \maketitle

%

\section{Introduction}

Most comets are expected to be remnants from the early times, some 4.5 billion years ago, when the Solar System was formed, and therefore they may contain records of the chemical and physical properties of these times past.  Exceptions to this notion
are, of course, the recent discoveries of the objects 1I/'Oumuamua and 2I/Borisov which are suggested to be interstellar visitors entering
our Solar System with strongly hyperbolical, and thus unbound, trajectories. When approaching the Sun to within a few au, comets can start to sublimate molecules
and other volatile particles from their surfaces and form what is known as a cometary coma. The study of the composition of cometary comae content can therefore
reveal information on the physical conditions and chemistry prevailing when the Solar System was born or similar systems in the solar neighbourhood in case of the interstellar
visitors.

The study of the cometary coma molecular content is performed either by remote sensing (mainly by recording spectral signatures over
a wide range of wavelengths from radio to the ultraviolet regimes) or by in-situ measurements made by spacecraft like Giotto (1P/Halley) and, more recently, Rosetta (67P/Churyumov-Gerasimenko). As pointed out by \citet{Rubin2019}, most abundances of the volatile content, determined by these various means over the last 40 yrs, are
reminiscent of those in the interstellar medium and thus suggest a chemical origin in interstellar, star-forming material.
The major gaseous constituent of the neutral inner coma is water molecules
which other species relate to. For instance, the next two most abundant species, CO$_2$ and CO,
constitute typically 10-20\% of that of water \citep{MummaCharnley2011, Rubin2019}. Further out in the coma, the
solar radiation will affect the composition. At 1 au from the Sun, the outgassing water molecules are photo-dissociated into
H and OH at a rate of about $(1-2)\times 10^{-5}\,\mathrm{s^{-1}}$ \citep{Huebner1992} where the factor of two variation reflects the degree of solar activity. An about 30 times slower water destruction mechanism is the photo-ionization into H$_2$O$^+$. This latter process, together with photo-ionization of OH, previously formed from water, is the main source of electrons in the coma \citep{Rubin2009}.
Since water is relatively difficult to spectroscopically observe, several other, indirect, methods have
been employed to determine the water production rate. Firstly, the observations of
the photo-dissociation product OH can be used as a proxy for water \citep{Despois1981, Bonev2006}. Secondly, also observations of the remaining
H, via Ly-$\alpha$, can be used as
exemplified by the Solar Wind ANisotropies (SWAN) Ly-$\alpha$ camera observations onboard the SOlar and Heliospheric Observer (SOHO) spacecraft \citep[eg.][]{Combi2011}. A third indirect method is to use a relatively abundant parent species, like HCN, as
an indicator of water -- see \citet{MummaCharnley2011} for an overview. Direct observations of the ground state water lines are scarce and are mostly limited to satellites \citep[e.g.][]{Lecacheux2003, Lis2013, Biver2015}. Infrared spectroscopy of the ro-vibrational water lines has proven useful to determine the water production rates \citep{Mumma2003}. This latter
method, albeit relying on fluorescence pumping models, has the advantage that the excitation conditions can be determined, via a rotation temperature, and consequently provide more accurate production rates  \citep[e.g.][]{Disanti2016}. Lastly, in-situ measurements by spacecraft, via high-resolution mass spectroscopy, provide perhaps the most direct way, although limited to a few comets, to probe the molecular coma content \citep[e.g.][]{Lauter2020}.

The observations presented in this paper are part of an ongoing effort to study comets at mm wavelengths using the Onsala Space
Observatory (OSO) 20-meter telescope. This effort was initiated by \citet{Wirstrom2016} who presented HCN(1-0) observations of
the long-period comets C/2013 R1 (Lovejoy) and C/2014 Q2 (Lovejoy). We here expand this study with observations of another twelve comets.
The observational efforts focus primarily on using HCN(1-0) as an indicator of water but one comet, 46P/Wirtanen, was also observed in methanol. In fact, both these molecules are believed to be released from nucleus ices \citep{DelloRusso2016}. Our comet target sample comprises bright comets with perihelion dates from late 2016 to 2019. The sample includes comets belonging to the Jupiter family as well as the Oort cloud. In addition, we also searched for HCN towards 2I/Borisov. Furthermore, we also investigate possible caveats, in terms of time-dependent radiative transfer effects, that may complicate the interpretation of the observed emission from HCN (as well as from other molecules) in cometary comae.

This paper is structured as follows. In the next section we describe the observations made with the OSO 20-m
telescope. In Sect.~\ref{sect:res} we present the results and then, in Sect.~\ref{sect:rad}, we outline the radiative transfer analysis made in
which we also take into account time-dependent aspects. After discussing our results in Sect.~\ref{sect:dis}, we finally make our conclusions.

\section{Observations}
\label{sect:obs}
All observations presented in this paper were made from late 2016 until late 2019 using the radome-enclosed OSO 20-m antenna equipped with the 3 mm channel of the 3 and 4~mm receiver system \citep{Belitsky2015}. The radome allows us to observe sources near the Sun without thermal distortion of the telescope optics.
The rest frequencies of the target lines, see Table~\ref{tab:lines},
were covered by two frequency tunings, one centred near 88 GHz, and the other near 96 GHz. All tunings were performed in the velocity
frame of the comet. Coordinates and velocities of all comets were taken from the Horizons system\footnote{{\it HORIZONS} is a service of the Solar System Dynamics group at the NASA/Jet Propulsion Laboratory for computing ephemerides of solar-system bodies. The web interface is at {\url{https://ssd.jpl.nasa.gov/horizons.cgi}}} \citep{Giorgini1996} tabulated at every 1 hr and then interpolated by the observing system. The antenna
half-power beam width is 40~arcsec at the HCN frequency and about 37 arcsec for the CH$_3$OH observations.

\begin{table*}
\caption{Observed lines}
\label{tab:lines}
\begin{tabular}{lcccrc}
\hline
\hline
Molecule & Frequency\tablefoottext{a} &  Transition          & $g_u$ & \multicolumn{1}{c}{$E_u$\tablefoottext{b}}   & $A_{ul}$\tablefoottext{c} \\
         & (MHz)     &                      &       & \multicolumn{1}{c}{(K)} &  (s$^{-1}$) \\
\hline
HCN      & 88630.42    & $J_F=1_1-0_1$ & 3   & 4.2 & $2.4\times 10^{-5}$ \\
         & 88631.85    & $J_F=1_2-0_1$ & 5   & 4.2 & $2.4\times 10^{-5}$ \\
         & 88633.94    & $J_F=1_0-0_1$ & 1   & 4.2 & $2.4\times 10^{-5}$ \\
\hline
CH$_3$OH & 95169.39    &  $J_K=8_{0}-7_{1}\, A^+$  & 17 & 83.5 & $4.3\times 10^{-6}$ \\
         & 95914.31    &  $J_K=2_{1}-1_{1}\, A^+$  & 5  & 21.4 & $2.5\times 10^{-6}$ \\
         & 96739.36    &  $J_K=2_{-1}-1_{-1}\, E$  & 5  & 12.5 & $2.6\times 10^{-6}$ \\
         & 96741.37    &  $J_K=2_{0}-1_{0}\, A^+$  & 5  &  7.0 & $3.4\times 10^{-6}$ \\
         & 96744.54    &  $J_K=2_{0}-1_{0}\, E$    & 5  & 20.1 & $3.4\times 10^{-6}$ \\
         & 96755.50    &  $J_K=2_{+1}-1_{+1}\, E$  & 5  & 28.0 & $2.6\times 10^{-6}$ \\
\hline
\end{tabular}
\tablefoot{
\tablefoottext{a}{Frequencies have been taken from \cite{Pickett1998}}
\tablefoottext{b}{Upper energy for methanol $E$-species includes the 7.9~K offset relative the $J_K=0_0$ level of the $A$-species}
\tablefoottext{c}{$A$-coefficients from \cite{Muller2001}}
}
\end{table*}

The Fast Fourier Transform Spectrometers (FFTSs) were configured into a wide 4-GHz mode or a narrow 156 MHz mode. In both modes, both linear polarizations
are recorded separately. The wide 4-GHz mode is made up of two partly overlapping 2.5 GHz FFTSs of 32768 channels and was used for the
CH$_3$OH observations and part of the HCN observations. The wide mode results in a velocity resolution of about 0.25~$\mathrm{km\, s^{-1}}$.
Normally, the narrow 156 MHz setup was used for the HCN observations. The observations were made in frequency switching mode with a
frequency throw for the HCN setup of 5~MHz for data taken in 2016-2017 and 7~MHz from 2018 and onwards. When observing CH$_3$OH, the wide spectrometer setup was employed to cover all lines, see Table~\ref{tab:lines}, and the
used frequency throw was larger, 25~MHz, to avoid line confusion.

The weather conditions during our observations varied. In the best conditions the system temperature was around 120~K.
Typically the measurements were automatically halted when the system temperature became large, around 800-1200~K. Pointing and focus optimizations,
using the SiO 2-1 $v=1$ maser line at 86 GHz towards stars, were performed several times a day.
 Measurements from 2018 and onwards were done with the new antenna control system,
developed by M. Lerner and called BIFROST\footnote{see \url{https://www.chalmers.se/en/researchinfrastructure/oso/radio-astronomy/20m/Pages/Handbook.aspx} and \url{http://www.ira.inaf.it/eratec/florence/presentations/florens_Lerner.pdf}}, which provides an
improved and more automated control of remote observations.
We have used these capabilities to perform pointings and focussings automatically every three hours and to automatically pause observations when the system temperature has gone above 800 K.
We have been using 60~s and 120~s long integrations for individual scans. With the switch to BIFROST we employed observing blocks consisting of fifteen 60-second scans with hot-load calibrations between every fifth scan.

Information on the 12 comets investigated in this study is summarized in Table~\ref{tab:sample} where the range of observing date, total integration time,
heliocentric distance, $R_\mathrm{h}$, and distance to Earth, $\Delta$, variations over the observing dates also have been entered. Also the projected nucleocentric distance (near the center of the date range), $d$, as subtended by the beam size for HCN(1-0) transition has been listed.

\begin{table*}
\caption{\label{tab:sample} Observed comets with the OSO 20-m telescope}
\centering
   \begin{tabular}{lccccc}
   \hline\hline
Comet               & Observing UT date range         & Int. time & $R_\mathrm{h}$ &   $\Delta$  & $d$ \\
                    &                                 &   (hr)    & (au)           &    (au)     & (km) \\
\hline
\object{C/2016 U1 (NEOWISE)} & 2016-12-22 -- 2017-01-10 & 14 & 0.71 -- 0.34  &  0.75 -- 1.07 & $5.2\times 10^4$ \\
\object{45P/Honda-Mrkos-Pajdu\u{s}\'{a}kov\'{a}} & 2017-01-18 --  2017-02-06 & 31 & 0.65 -- 0.90 & 0.36 -- 0.11 & $7.0\times 10^3$ \\
\object{2P/Encke}            &  2017-03-04 -- 2017-03-17 & 15 & 0.37 -- 0.38 &  0.75 -- 0.68 & $4.1\times 10^4$ \\
\object{41P/Tuttle-Giacobini-Kres\'{a}k} &  2017-04-04 -- 2017-04-24 & 170 & 1.05 -- 1.06 & 0.14 -- 0.17 & $8.7\times 10^3$ \\
\object{C/2015 ER61 (PanSTARRS)} &  2017-04-06 -- 2017-05-09 & 12 & 1.19 -- 1.04 & 1.21 -- 1.23 & $7.1\times 10^4$ \\
\object{C/2017 E4 (Lovejoy)} &  2017-04-07                  & 8 & 0.63 & 0.67 & $3.9\times 10^3$ \\
\object{C/2015 V2 (Johnson)} &  2017-05-18  -- 2017-05-19   & 13 & 1.67 & 0.87 & $5.0\times 10^4$ \\
\object{C/2017 O1 (ASASSN1)} & 2017-10-12 -- 2017-10-15  & 30 & 1.50 & 0.73 -- 0.72 & $4.2\times 10^4$ \\
\object{96P/Machholz}        & 2017-10-30                  & 5 & 0.16 & 1.04 & $6.0\times 10^4$ \\
\object{46P/Wirtanen}        & 2018-12-08 -- 2019-01-18 & 232 & 1.06 -- 1.16 & 0.096 -- 0.22 & $4.7\times 10^3$ \\
\object{C/2018 Y1 (Iwamoto)} & 2019-02-04 -- 2019-02-19 & 38 & 1.29 -- 1.30 & 0.46  -- 0.39 & $2.4\times 10^4$ \\
\object{2I/Borisov}          & 2019-10-28 -- 2019-11-18 & 89 & 2.21 -- 2.06 & 2.66  -- 2.18 & $1.4\times 10^5$ \\
 \hline
    \end{tabular}
\tablefoot{Towards 96P/Machholz only CH$_3$OH was searched for. Both HCN and CH$_3$OH were observed in 46P/Wirtanen. For the rest of the comets only HCN was targeted.}
\end{table*}

\section{Results}
\label{sect:res}
\begin{figure*}
\resizebox{\hsize}{!}{
\includegraphics{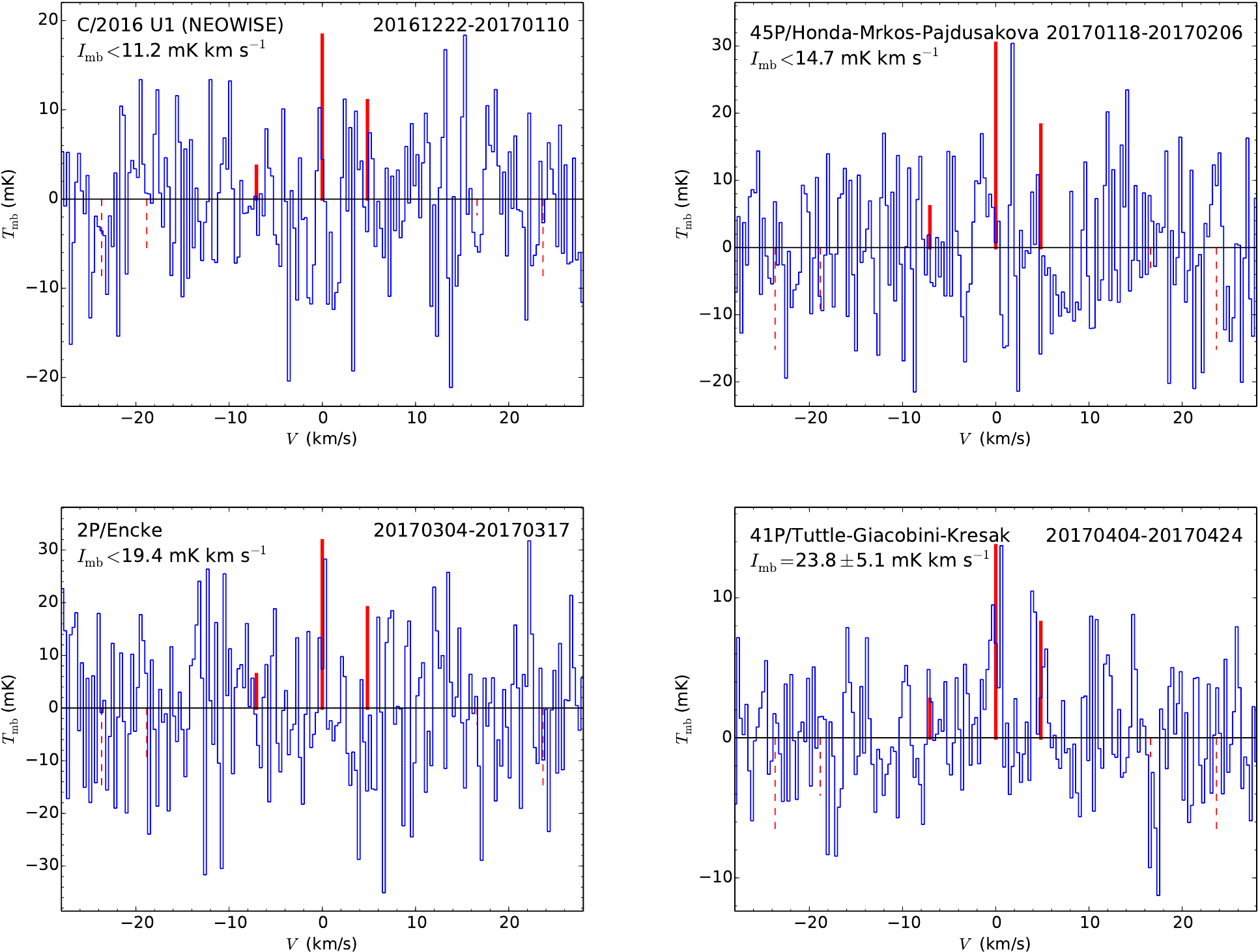}
}
\caption{OSO 20-m HCN 1-0 spectra towards the comets U1, 45P, 2P/Encke, and 41P. The velocity scale is in the
reference frame of the comets and the intensity scale is in $T_\mathrm{mb}$. The positions of the three hfs components are marked with red
bars. The relative heights of the bars correspond to the statistical weights, $g_u$, in Table~\ref{tab:lines}.
Also the integrated intensity is shown together with $1\sigma$ error or as $1\sigma$ upper limit where
applicable.
}
\label{fig:spe1}
\end{figure*}

\begin{figure*}
\resizebox{\hsize}{!}{
\includegraphics{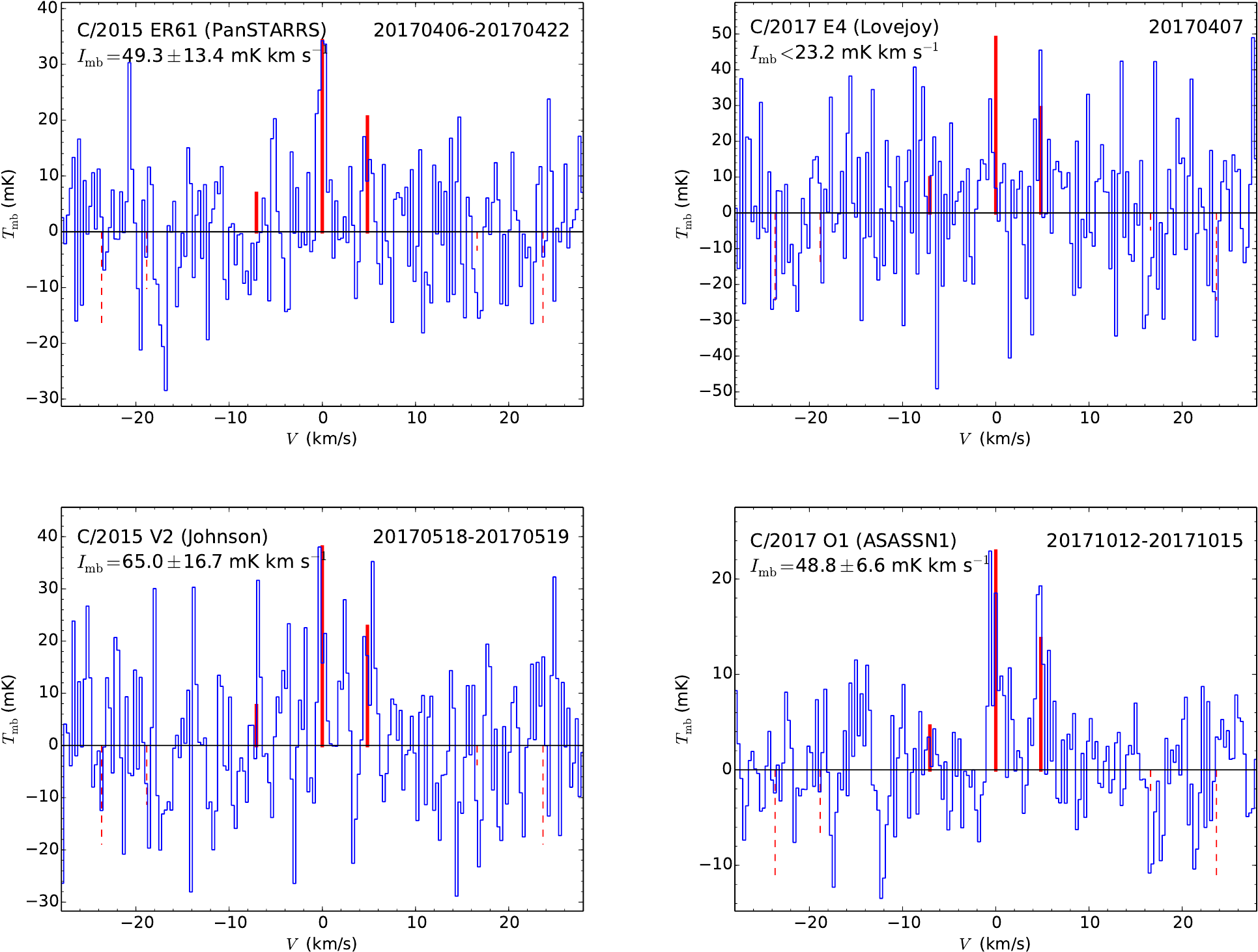}
}
\caption{HCN 1-0 spectra towards the comet ER61, E4, V2, and O1. Scales as in Fig.~\ref{fig:spe1}}
\label{fig:spe2}
\end{figure*}

\begin{figure}
\resizebox{\hsize}{!}{
\includegraphics{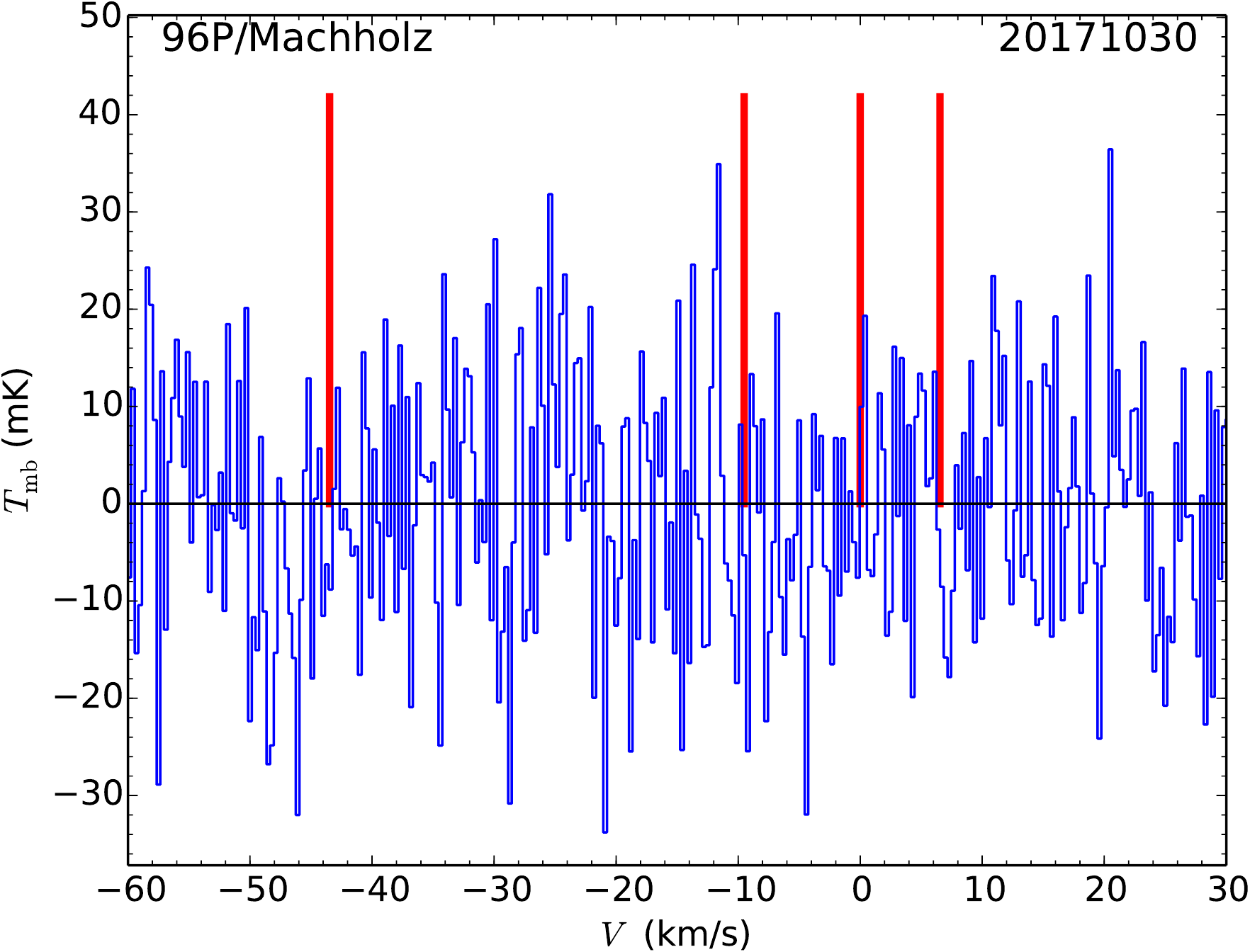}
}
\caption{CH$_3$OH spectrum towards the comet 96P/Machholz on 30 October 2017. The positions where the CH$_3$OH lines would
appear are indicated by the red bars. Scales as in Fig.~\ref{fig:spe1}}
\label{fig:machholz}
\end{figure}

\begin{figure}
\resizebox{\hsize}{!}{
\includegraphics{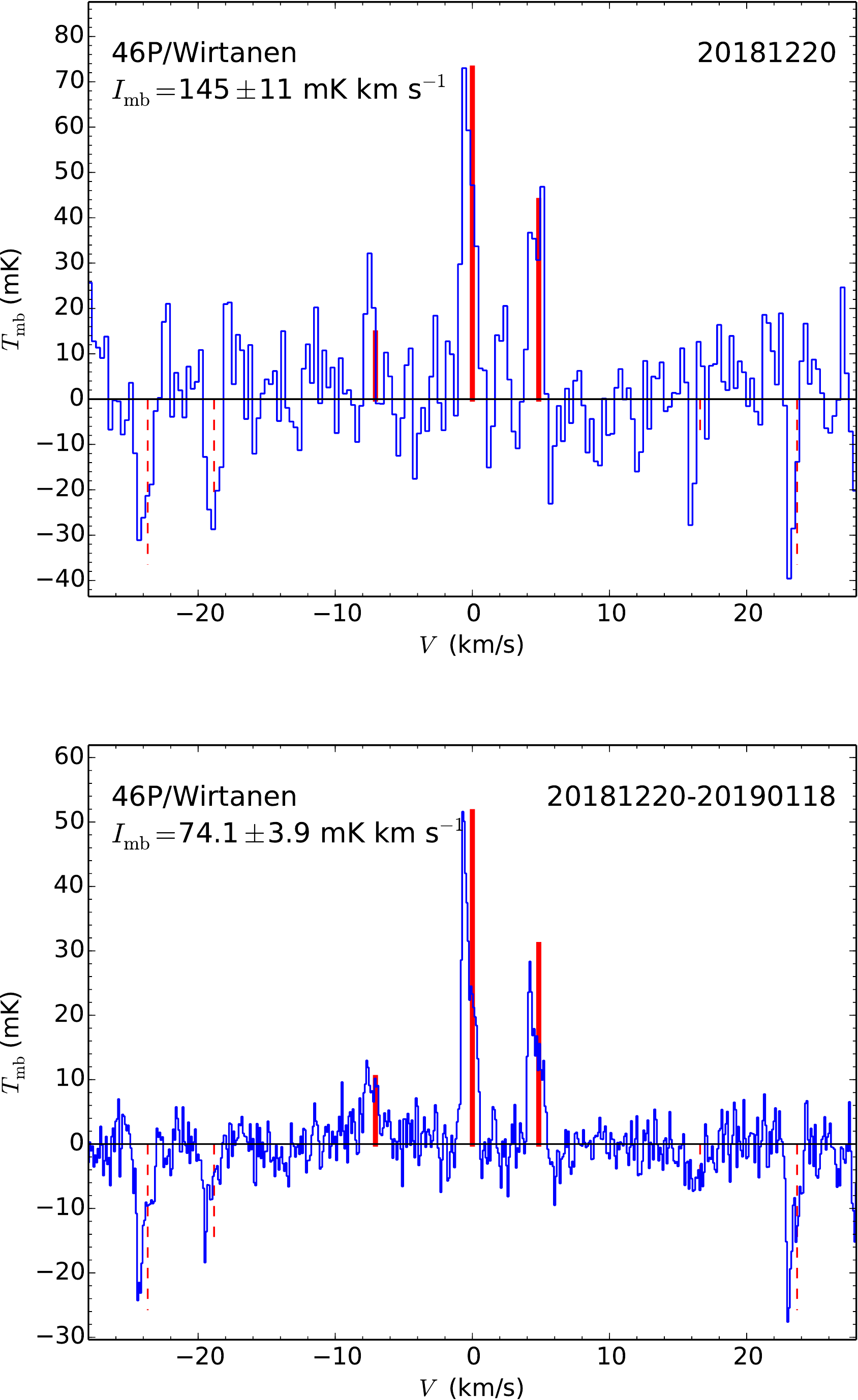}
}
\caption{Top: HCN 1-0 spectrum towards the comet 46P/Wirtanen on 20 December 2018. The positions, corresponding to a frequency throw of
7 MHz, of the negative artifacts of the HCN lines have been marked with negative dashed red lines.
Bottom: Global average of all high-resolution ($0.1\,\mathrm{km\, s^{-1}}$) HCN data observed from 20 December 2018 to 18 January 2019.
Scales as in Fig.~\ref{fig:spe1}.
}
\label{fig:wirtanen}
\end{figure}

\begin{figure*}
\sidecaption
\includegraphics[width=12cm]{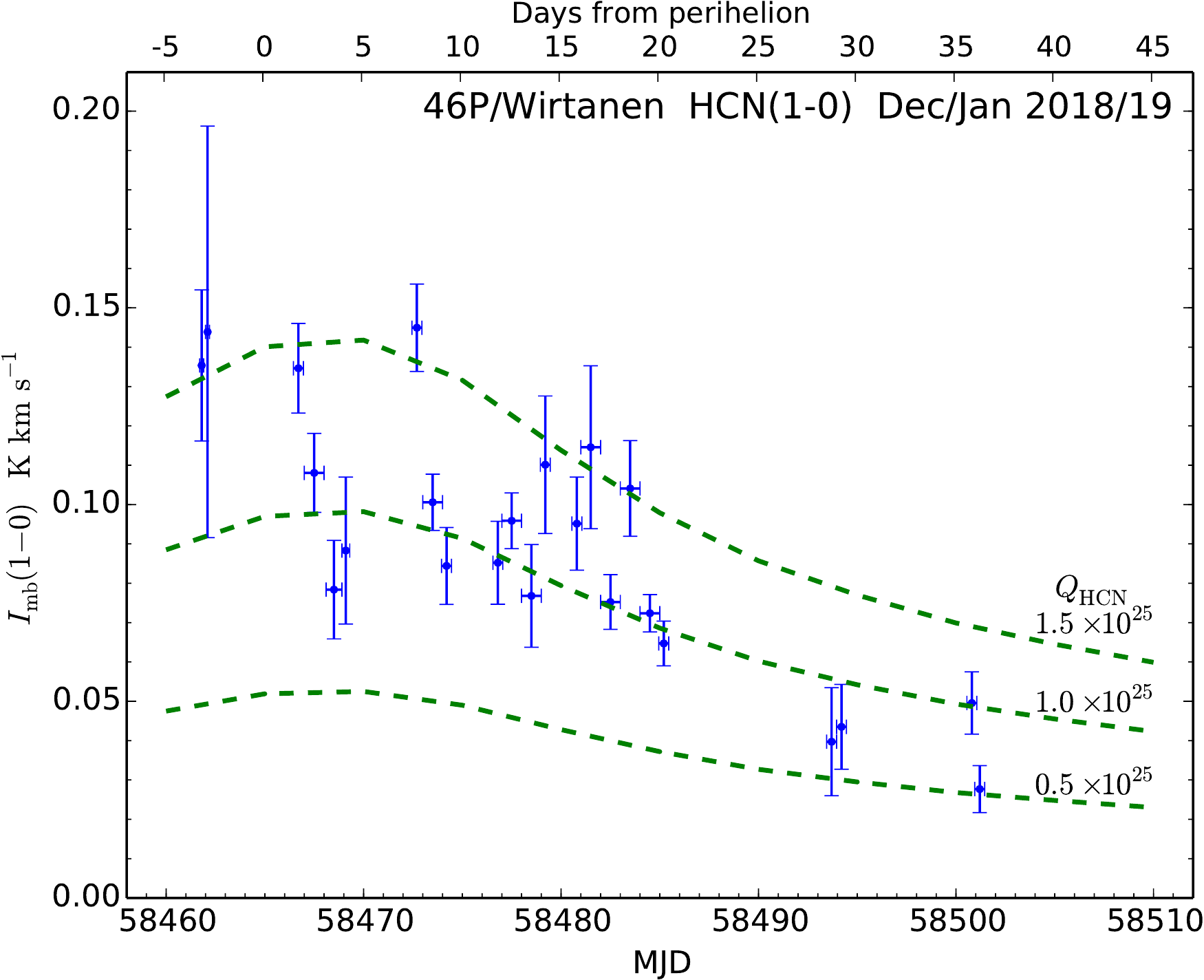}
\caption{OSO 20-m HCN 1-0 monitoring data towards 46P/Wirtanen in December 2018 and January 2019. The data points
(blue marker with vertical $1\sigma$ errors bars, horizontal bars indicate range of observations), see Table~\ref{tab:wirtanen_results}, represent the integrated intensity as
a function of modified Julian date (MJD) or days from perihelion. The dashed lines represent the results from the radiative transfer modelling, see Sect.~\ref{sect:rad}, using
3 different HCN production rates from $0.5\times 10^{25}\,\mathrm{mol\, s^{-1}}$ to $1.5\times 10^{25}\,\mathrm{mol\, s^{-1}}$ as indicated for a temperature of 70~K and $Q_\mathrm{HCN}/Q_\mathrm{H_2O} = 0.1\mathrm{\%}$.}
\label{fig:wirt_hcn}
\end{figure*}

\begin{figure}
\resizebox{\hsize}{!}{
\includegraphics{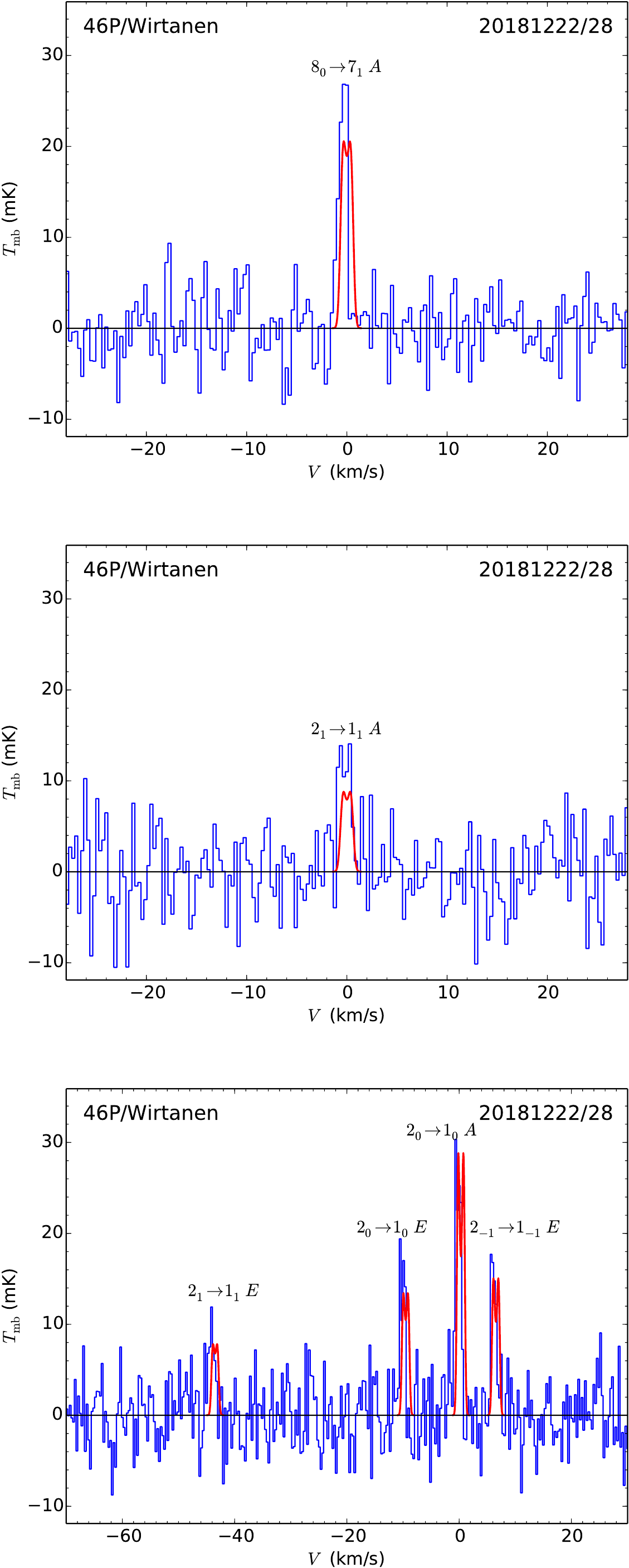}
}
\caption{OSO 20-m methanol spectra towards 46P/Wirtanen when averaging data between 22 and 28 December 2018. The red spectra represent the modelling results of the radiative transfer modelling, see Sect.~\ref{sect:rad}, using
a CH$_3$OH production rate of $1.6\times 10^{26}\,\mathrm{mol\, s^{-1}}$ and a kinetic temperature of 70~K.}
\label{fig:wirt_ch3oh}
\end{figure}

\begin{figure}
\resizebox{\hsize}{!}{
\includegraphics{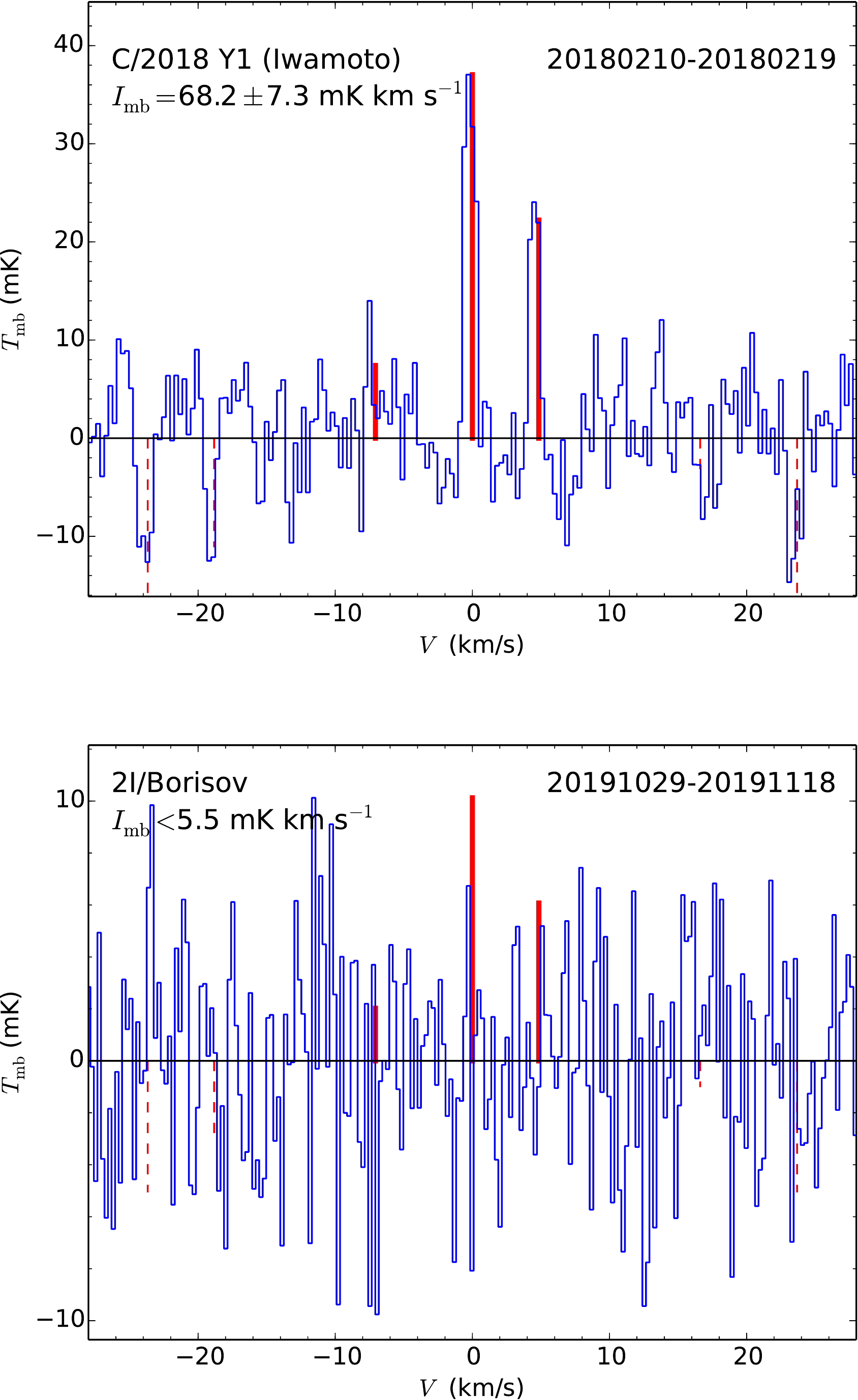}
}
\caption{Top: OSO 20-m HCN 1-0 spectrum towards C/2018 Y1 (Iwamoto). It is an average of data between 10 and 19 February 2019.
Bottom: The HCN spectrum observed towards 2I/Borisov in October/November 2019. Scales as in Fig.~\ref{fig:spe1}.
}
\label{fig:spe3}
\end{figure}

The HCN and CH$_3$OH observations are displayed in Figs. \ref{fig:spe1}-\ref{fig:spe3}. The shown spectra have been smoothed to a resolution of
$0.3\,\mathrm{km\, s^{-1}}$ (except the 46P/Wirtanen global average spectrum) and intensities are shown in $T_\mathrm{mb}$-scale. The individual scans were first folded before being noise-weighted to form an average. Because of
the frequency switching observing mode, a rather high order polynomial was needed to subtract the baseline. The used baseline order was typically from 3 to 7 where the higher orders were used for the observations with larger frequency throws. The channel RMS and integrated line
intensities, $I_\mathrm{mb}$, have been entered in Table~\ref{tab:hcn_results}. Errors and upper limits are $1\sigma$
in this table. The variation of errors and limits among the comets is a combination of system temperature and integration time.
The integrated line intensity of HCN($J=1\to 0$) is calculated as a sum of the three hyperfine structure (hfs) line intensities, listed in Table~\ref{tab:lines},
 using a box of width $2\,\mathrm{km\, s^{-1}}$ centered on each hfs line. The 46P/Wirtanen monitoring measurements have
 been entered in Table~\ref{tab:wirtanen_results}. Below we give a short introduction and summary of the results for each
 comet observed:

\begin{itemize}

\item{C/2016 U1 (NEOWISE)}. This comet (hereafter U1), discovered in October 2016 by the Near-Earth Object Wide-field Infrared Survey Explorer
(NEOWISE), follows a slightly hyperbolic orbit (eccentricity just above 1). At perihelion on 14 January 2017 its heliocentric distance was 0.32~au. The spectrum presented in
Fig.~\ref{fig:spe1} is the average over 4 observing dates ranging from 22 December 2016 to 10 January 2017. It was not detected in HCN and the
upper limit of the integrated intensity is listed in Table~\ref{tab:hcn_results}.

\item{45P/Honda-Mrkos-Pajdu\u{s}\'{a}kov\'{a}}. This short-period comet discovered in 1948, hereafter called 45P, belongs to the Jupiter-family and has a period of 5.26 yr. The water production was estimated by \citet{DelloRusso2020} to be
$(2-3)\times 10^{27}\,\mathrm{mol\, s^{-1}}$ using infrared spectroscopy in mid-February 2017. The same study reports an HCN production rate of $(3-4)\times 10^{24}\,\mathrm{mol\, s^{-1}}$. On 16 February 2017 there was
also a CN outburst seen \citep{Springmann2019}. The HCN 1-0 triplet was not detected by us, see Fig.~\ref{fig:spe1}. The bulk of our HCN data was obtained early February 2017.

\item{2P/Encke}. This short-period comet was observed several times in the first half of March 2017. Our HCN observations resulted in a non-detection. Later in March, after perihelion, the water production rate was estimated to be $(3-4)\times 10^{28}\,\mathrm{mol\, s^{-1}}$ and the
HCN production rate was $(3-6)\times 10^{25}\,\mathrm{mol\, s^{-1}}$ \citep{Roth2018}.

\item{41P/Tuttle-Giacobini-Kres\'{a}k}. It was first observed in 1858 and later rediscovered in 1907 and 1952 as
pointed out by \citet{Schleicher2019}. Like 45P, this comet (hereafter called 41P) belongs to the Jupiter-family and has a period of 5.42 yr.
In early April 2017 its peak water production rate was $3.5\times 10^{27}\,\mathrm{mol\, s^{-1}}$ \citep{Moulane2018}. The 41P nucleus rotation did undergo an unusually large slow-down with a rotation period increasing from 20 hr in March 2017 to more than 50 hr in early May \citep{Schleicher2019}.
Our April 2017 HCN 1-0 detection, at the 4.6$\sigma$ level, is shown in Fig.~\ref{fig:spe1}.

\item{C/2015 ER61 (PanSTARRS)}. This comet, hereafter ER61, follows a highly eccentric orbit and is suggested to originate in the inner Oort cloud \citep{Meech2017}. The OSO 20-m HCN 3.7$\sigma$ detection is shown in Fig.~\ref{fig:spe2}. The spectrum is an average over three observing days
in 2017; 6, 11, and 22 April. The observations took place after ER61 underwent an outburst on 4 April \citep{Opitom2019}. \citet{Sekanina2017} suggests that the flare-up was caused by fragmentation of
the nucleus creating a companion nucleus which was observed later in June. Our observations on 9 May were disregarded due to an inaccurate set of ephemeris. \citet{Saki2021} report a water production rate
around $1\times 10^{29}\,\mathrm{mol\, s^{-1}}$ in mid-april 2017 and an HCN rotational temperature near 70~K. This is consistent with the Atacama Large Millimeter/Submillimeter Array (ALMA) compact array HCN observations in mid-April by \citet{Roth2021c}. Both studies determined the HCN production rate to be near $9\times 10^{25}\,\mathrm{mol\, s^{-1}}$.

\item{C/2017 E4 (Lovejoy)}. This long-period comet (referred to as E4) appears to have originated in the inner Oort Cloud \citep{Faggi2018}. Our observations took place on 7 April and resulted in an upper limit, see Fig.~\ref{fig:spe2}. \citet{Faggi2018} reported a water production rate of about
$3\times 10^{28}\,\mathrm{mol\, s^{-1}}$ and an HCN production rate of about $5\times 10^{25}\,\mathrm{mol\, s^{-1}}$ only 3 days before our observations took place. The comet disintegrated in late April 2017.

\item{C/2015 V2 (Johnson)}. This Oort cloud comet (OCC), hereafter referred to as V2, follows a slightly hyperbolic orbit and was observed during one observing run stretching from 18-19 May in 2017 before the perihelion on 12 June. These observations resulted in an HCN 3.9$\sigma$ detection, see Fig.~\ref{fig:spe2}. \citet{Combi2021} have estimated water production rates from SOHO/SWAN observations but no other direct
molecular studies of V2 during this period of time have been found in the literature.

\item{C/2017 O1 (ASASSN1)}. This long-period comet, hereafter called O1, is a possible Manx comet \citep{Brinkman2020}, i.e., a category of tail-less long-period comets which may have formed in the inner solar system and then been ejected to the outskirts. Our observation, resulting in a 7.4$\sigma$ detection (Fig.~\ref{fig:spe2}), took place in two shifts in mid October 2017 (12/13 and 14/15).

\item{96P/Machholz}. This short-period comet has a rather peculiar high-inclination, low-perihelion orbit being a Jupiter family comet (JFC). Its
nucleus is thought to be inactive \citep[e.g.][]{Eisner2019}. Our observations, of CH$_3$OH only, were done
when 96P/Machholz was at a heliocentric distance of 0.16 au on 30 October 2017 (the perihelion distance
was 0.12 au for the 2017 apparition). Emission from CH$_3$OH was not detected with the OSO 20-m at this date,
see Fig. ~\ref{fig:machholz}, and the upper limit for the $J_K=2_{0}-1_{0}\, A^+$ line has been included in
Table~\ref{tab:ch3oh_results}.

\item{46P/Wirtanen}. This JFC was the only comet in our sample that could be monitored in HCN, see
Fig.~\ref{fig:wirt_hcn} and Table~\ref{tab:wirtanen_results}. The first detection of HCN was made on 9 December 2018 and the last detection was made on 18 January 2019. The detection level was better than 3$\sigma$ except
on the dates 10 December and 10 January.
On 20 December we detected the strongest HCN line intensity. The HCN spectrum of this date is shown in Fig.~\ref{fig:wirtanen}. Also CH$_3$OH was clearly detected, the spectrum (an average of data taken between 22 and 28 December) is shown in Fig.~\ref{fig:wirt_ch3oh}. Six different methanol lines around 96 GHz are seen (Table~\ref{tab:ch3oh_results}), each at a detection level of 5$\sigma$ or better. It came very close to
Earth in December 2018 (0.08 au) and was hence widely observed\footnote{\url{https://wirtanen.astro.umd.edu/46P/46P_2018.shtml}}.
In addition, it showed much higher activity than during previous apparitions \citep{Farnham2019}.
HCN was observed by \citet{Wang2020} ($J=1-0$) and \citet{Coulson2020} ($J=4-3$). The latter study also included $J=7-6$ CH$_3$OH data. A week before closest approach to Earth,
\citet{Roth2021b} observed the $J=5-4$ CH$_3$OH lines using the ALMA array. They found variable CH$_3$OH outgassing
consistent with the rotational period of 9 hrs \citep{Farnham2021} of the nucleus. In December 2018
\citet{Biver2021} performed a molecular survey of 46P mainly using the Institut de RadioAstronomie Millim{\'e}trique (IRAM) 30-m telescope but also the
NOrthern Extended Millimeter Array (NOEMA).
At perihelion (12.9 December 2018) \citet{Moulane2019} reported a water production rate of $7.2\times 10^{27}\,\mathrm{mol\, s^{-1}}$ followed by a number of other infrared spectroscopy studies \citep[][]{Saki2020, Roth2021a, Bonev2021, Khan2021, McKay2021} as well as submillimetre observations of $\mathrm{H_2^{18}O}$ \citep{Lis2019}.

\item{C/2018 Y1 (Iwamoto)}. This bright and long-period comet, hereafter called Y1, came close (0.4~au) to Earth in February 2018. On February 5 \citet{Disanti2019} report H$_2$O and HCN production rates of
$2\times 10^{28}\,\mathrm{mol\, s^{-1}}$ and $4\times 10^{25}\,\mathrm{mol\, s^{-1}}$, respectively.
By averaging data between 10 and 19 February we obtained a very clear 9$\sigma$ detection of HCN, see Fig.~\ref{fig:spe3}.
Data taken on February 4 were excluded in the analysis because of erroneous Doppler corrections.

\item{2I/Borisov}. Being only the second interstellar object known to have visited our Solar System, it was the first such object to
show outgassing activity. This was confirmed already in September 2019 \citep{Fitzsimmons2019}. From interferometry imaging observations, using the ALMA \citep{Cordiner2020}, and Hubble Space Telescope (HST) observations
\citep{Bodewits2020}, a CO production rate of about $(5-10)\times 10^{26}\,\mathrm{mol\, s^{-1}}$ around perihelion was found.
\citet{Cordiner2020} also determined the HCN production rate to be $7\times 10^{23}\,\mathrm{mol\, s^{-1}}$. The 20-m observations in
October and November 2018 only resulted, when averaging all data, in an upper limit of the HCN 1-0 emission, see Fig.~\ref{fig:spe3}.

\end{itemize}

\begin{table}
\caption{\label{tab:hcn_results} Observed HCN intensities and upper limits}
\centering
   \begin{tabular}{lrc}
   \hline\hline
Comet                                   & Channel rms & $I_\mathrm{mb}$ \\
                                        & (mK) & ($\mathrm{mK\, km\, s^{-1}}$) \\
\hline
C/2016 U1 (NEOWISE)                     &  7.5  & $<11.2$ \\
45P/Honda-Mrkos-Pajdu\u{s}\'{a}kov\'{a} &  9.8  & $<14.7$ \\
2P/Encke                                & 12.9  & $<19.4$ \\
41P/Tuttle-Giacobini-Kres\'{a}k         &  3.8  & $23.8 \pm 5.1$ \\
C/2015 ER61 (PanSTARRS)                 & 10.0  & $49.3\pm 13.4$ \\
C/2017 E4 (Lovejoy)                     & 17.3  & $<23.2$ \\
C/2015 V2 (Johnson)                     & 12.5  & $65.0\pm 16.7$ \\
C/2017 O1 (ASASSN1)                     &  5.0  & $48.8\pm 6.6$ \\
C/2018 Y1 (Iwamoto)                     &  5.5  & $68.2 \pm 7.3$\\
2I/Borisov                              &  4.0  & $<5.5$ \\
 \hline
    \end{tabular}
\end{table}

\begin{table}
\caption{\label{tab:wirtanen_results} 46P/Wirtanen HCN 2018/19 monitoring results}
\centering
   \begin{tabular}{lccccc}
   \hline\hline
Date      & MJD   &   $R_\mathrm{h}$  & $\Delta$  & $I_\mathrm{mb}$  & $Q_\mathrm{HCN}/10^{25}$ \\
          &       &  (au)  & (au)      & $\mathrm{mK\, km/s}$ & ($\mathrm{mol/s}$)\\
\hline
1209   & 58461.8 &  1.056 &  0.088   & $135\pm 19$  & $1.57\pm 0.21$ \\
1210   & 58462.1 &  1.056 &  0.086   & $144\pm 52$  & $1.65\pm 0.60$ \\
1214   & 58466.7 &  1.056 &  0.078   & $135\pm 11$  & $1.41\pm 0.11$ \\
1215   & 58467.5 &  1.056 &  0.078   & $108\pm 10$  & $1.10\pm 0.10$ \\
1216   & 58468.5 &  1.056 &  0.077   &  $78\pm 13$  & $0.76\pm 0.12$ \\
1217   & 58469.1 &  1.057 &  0.078   &  $88\pm 19$  & $0.88\pm 0.18$ \\
1220   & 58472.7 &  1.060 &  0.081   & $145\pm 11$  & $1.57\pm 0.12$ \\
1221   & 58473.5 &  1.062 &  0.083   & $101\pm  7$  & $1.08\pm 0.07$ \\
1222   & 58474.2 &  1.063 &  0.085   &  $84\pm 10$  & $0.89\pm 0.10$ \\
1224   & 58476.8 &  1.067 &  0.091   &  $85\pm 11$  & $0.96\pm 0.12$ \\
1225   & 58477.5 &  1.069 &  0.095   &  $96\pm  7$  & $1.14\pm 0.08$ \\
1226   & 58478.5 &  1.071 &  0.098   &  $77\pm 13$  & $0.91\pm 0.15$ \\
1227   & 58479.2 &  1.074 &  0.102   & $110\pm 17$  & $1.41\pm 0.21$ \\
1228   & 58480.8 &  1.076 &  0.106   &  $77\pm 13$  & $0.97\pm 0.16$ \\
1229   & 58481.5 &  1.079 &  0.111   & $115\pm 21$  & $1.59\pm 0.28$ \\
1230   & 58482.5 &  1.082 &  0.115   &  $75\pm  7$  & $1.01\pm 0.09$ \\
1231   & 58483.5 &  1.085 &  0.120   & $104\pm 12$  & $1.51\pm 0.17$ \\
0101   & 58484.5 &  1.088 &  0.125   &  $73\pm  5$  & $1.05\pm 0.07$ \\
0102   & 58485.2 &  1.091 &  0.130   &  $65\pm  6$  & $0.95\pm 0.07$ \\
0110   & 58493.7 &  1.123 &  0.173   &  $40\pm 14$  & $0.67\pm 0.23$ \\
0111   & 58494.2 &  1.128 &  0.179   &  $44\pm 11$  & $0.77\pm 0.19$ \\
0117   & 58500.8 &  1.158 &  0.215   &  $50\pm  8$  & $1.02\pm 0.16$ \\
0118   & 58501.2 &  1.164 &  0.222   &  $28\pm  6$  & $0.52\pm 0.11$ \\
 \hline
    \end{tabular}
\tablefoot{Date is UT start of observation 2018/19mmdd. MJD corresponds to centre of time interval.
See Sect.~\ref{sec:modelresults} for the HCN production rates.
}
\end{table}

\begin{table}
\caption{\label{tab:ch3oh_results} Observed CH$_3$OH intensities and upper limit}
\centering
   \begin{tabular}{lcc}
   \hline\hline
Comet                                   & Line & $I_\mathrm{mb}$ \\
                                        &      & $\mathrm{mK\, km\, s^{-1}}$ \\
\hline
96P/Machholz        & $J_K=2_{0}-1_{0}\, A^+$ & $< 10.7$  \\
46P/Wirtanen        & $J_K=2_{0}-1_{0}\, A^+$ & $34.3\pm 2.7$ \\
                    & $J_K=2_{1}-1_{1}\, A^+$ & $20.0\pm 3.1$ \\
                    & $J_K=8_{0}-7_{1}\, A^+$ & $28.3\pm 2.7$ \\
                    & $J_K=2_{-1}-1_{-1}\, E$ & $19.6\pm 2.7$ \\
                    & $J_K=2_{0}-1_{0}\, E$   & $19.0\pm 2.7$ \\
                    & $J_K=2_{+1}-1_{+1}\, E$ & $13.4\pm 2.7$ \\
\hline
    \end{tabular}
\end{table}


In the case of 46P/Wirtanen (both HCN and CH$_3$OH) and C/2018 Y1 (Iwamoto) we see slightly shifted profiles in the
sense that their peaks appear shifted with $\sim 0.3-0.4\,\mathrm{km\, s^{-1}}$ to lower velocities, see
Figures~\ref{fig:wirtanen}, \ref{fig:wirt_ch3oh} and \ref{fig:spe3}.
This apparent blue-shift likely indicates that the gas outflow component directed
towards Earth is more pronounced. Since both these comets were at $R_\mathrm{h} > 1\,\mathrm{au}$ at the time of
observations, much of their sunward sides were then facing Earth. To illustrate this asymmetry more clearly a global
average of all HCN high-resolution data for 46P/Wirtanen is also
shown in Fig.~\ref{fig:wirtanen} for which the blue peak is centered at $-0.4\pm 0.1\,\mathrm{km\, s^{-1}}$ when fitted
with a Gaussian. The ratio of the blue to red wing emission for this spectrum is $2.6\pm 0.2$ when using the
velocity ranges $\pm (0.15-0.85)\,\mathrm{km\, s^{-1}}$ for the strongest $J_F=1_2-0_1$ line.
The observed line widths for 46P and Y1 seem to indicate an expansion velocity between $0.5$ and $0.6\,\mathrm{km\, s^{-1}}$.

\section{Radiative transfer modelling of comets}
\label{sect:rad}
As already mentioned, water is the main ingredient in the volatile material that is being outgassed from the comet surface. The gaseous material leaves the
comet surface with a nearly constant expansion velocity $v_\mathrm{e}$.
Further out in the coma, the water molecules may be photo-ionized into $\mathrm{H_2O^+}$ and photo-dissociated mainly into OH and H \citep{Rubin2009}. Other molecules, like HCN, behave in a similar way. In addition,
HCN is an important precursor to CN \citep{Fray2005, Paganini2010} although heated CN-bearing dust could be a significant source to gaseous CN according to \citet{Hanni2020}. This pure gas expansion may of course be modified when, for example, ions start to deflect when interacting with the Solar wind. This happens near what is called the contact surface, $R_\mathrm{CS}$ \citep{Rubin2009}.
However, the radial distribution of neutral parent molecules (i.e. those with origin from the comet surface) is thought to follow the Haser relation
\citep{Haser1957}
\begin{equation}
n_\mathrm{mol}(r) = \frac{Q_\mathrm{mol}}{4\pi r^2 v_\mathrm{e}}\, \exp\left(-r \gamma_\mathrm{p} / v_\mathrm{e} \right),
\label{eq:Haser}
\end{equation}
as a function of cometocentric radius $r$, where $Q_\mathrm{mol}$ is the molecular production rate and $\gamma_\mathrm{p}$ is the photo-dissociation rate at the
distance to the Sun.

\subsection{Basic equations}

As pointed out by \citet{Crovisier1983} the change rate of the physical conditions, like density decline due to
expansion, in the cometary comae
is not that different from the scale of molecular (in this case CO) radiative and collisional rates although their relative
importance vary with radius.
We here mainly follow the approach by \citet{Chin1984, BM1984, Crovisier1987, BM1987} when setting up the basic equations
governing the excitation.

For simplicity we start with a two-level system with upper level population $n_u$ and lower level population $n_l$. The total population is $n = n_u + n_l$.
The population change for the two levels can be expressed as the differential equations (DEs)
\begin{eqnarray}
 d n_u/dt &=& -n_u \left( A_{ul} + B_{ul}J_\nu + c_{ul} + \gamma_{u}\right) + n_l \left( B_{lu} J_\nu + c_{lu} \right) \label{eq:DE1} \\
 d n_l/dt &=& +n_u \left( A_{ul} + B_{ul}J_\nu + c_{ul}\right) - n_l \left( B_{lu} J_\nu + c_{lu} + \gamma_{l} \right),
\label{eq:DE2}
\end{eqnarray}
where $A_{ul}$ is the spontaneous decay from $u$ to $l$ and $B_{ul} J_\nu$ and $B_{lu} J_\nu$ are the rates due to stimulated emission and
absorption, respectively, when exposed to the averaged radiation field $J_\nu$ at the frequency $\nu$. The downward and upward collisional rates are
denoted $c_{ul}$ and $c_{lu}$, respectively. These are the standard processes included for a two-level system. We have also added two other destruction
rates referred to as $\gamma_u$ and $\gamma_l$. These could, for example, represent destruction due to photo-dissociation or photo-ionization. We will hereafter assume that the
destruction rate is equal for the upper and lower levels, that is, we set $\gamma = \gamma_u = \gamma_l$. By adding Eqs.~(\ref{eq:DE1}) and (\ref{eq:DE2}),
we then obtain
\begin{equation}
dn/dt = -\gamma\, n,
\label{eq:dn}
\end{equation}
which simply tells us that the total population $n(t)$ is exponentially decaying with time (if $\gamma > 0$).

In the case of a cometary coma, due to the outgassing, the molecules expand with a velocity according to $v_\mathrm{e}$, which we here assume is constant. The
cometocentric radius, $r$, is then given as $r = r_\mathrm{c} + v_\mathrm{e} t$ as a function of time. The radius of
the comet nucleus is $r_\mathrm{c}$. The destruction rate can then be modified to
\begin{equation}
\gamma = \gamma_\mathrm{p} + \frac{2}{r}\, dr/dt = \gamma_\mathrm{p} + \frac{2 v_\mathrm{e}}{r_\mathrm{c} + v_\mathrm{e} t},
\label{eq:gamma}
\end{equation}
where $\gamma_\mathrm{p}$ is the destruction rate due to photo-dissociation (or photo-ionization) and the second term, $2v_e/r$, reflects the population dilution due to constant expansion. We note that
the first part in Eq.(\ref{eq:gamma}) allows an expansion velocity varying with radius. However, we will hereafter always assume a constant gas expansion velocity.
By changing $dt$ to $dr/v_e$ in Eqs.~(\ref{eq:DE1}) and (\ref{eq:DE2}) it is straightforward to rewrite the population changes as a function of $r$ instead of $t$. For instance,
Eq.~(\ref{eq:dn}) then becomes
\begin{equation}
dn/dr = - (\gamma_\mathrm{p}/v_\mathrm{e} + 2/r)\, n,
\label{eq:dndr}
\end{equation}
which solves into the Haser distribution, Eq.~(\ref{eq:Haser}), when integrated. Hence, by adopting this approach we can study possible time-dependent effects of
the individual level populations $n_u$ and $n_l$ as function of cometocentric radius by solving the DEs in Eqs.~(\ref{eq:DE1}) and (\ref{eq:DE2}) while still
maintaining an overall decline due to the Haser equation. This may be of particular interest near the contact surface ($R_\mathrm{CS}$) where the electron
properties vary very rapidly (they enter into the equations via $c_{ul}$ and $c_{lu}$).

The above two-level system of equations is easily generalized into a multi-level system, but in order to simplify the solution of the DEs we here neglect
the contribution from the line itself when calculating the averaged radiation field, $\bar{J}_\nu$ at any given radial point in the coma. We assume that the
contributions to $\bar{J}_\nu$ only come from the diluted black body radiations of the comet nuclei, the Sun, and the 2.7~K cosmic background, respectively.
Here we use the geometric dilution factors for the comet nuclei, $\beta_\mathrm{c}(r)$, and that for the Sun, $\beta_\sun$. The remaining fraction of the
$4\pi$ sky background is assumed to be filled by the $T_\mathrm{cmb} = 2.7\,\mathrm{K}$ cosmic background radiation. Neglecting effects due to
shadowing, this can be expressed as
\begin{equation}
\bar{J}_\nu = \beta_\mathrm{c}(r)\, J(T_\mathrm{c}) + \beta_\sun\, J(T_\sun) + \left[1 - \beta_\mathrm{c}(r) - \beta_\sun \right] J(T_\mathrm{cmb}).
\label{eq:J}
\end{equation}
Here $J(T) = 1/(\exp(h\nu/kT)-1)$ where $h$ and $k$ are Planck's and Boltzmann's constants, respectively. The radiation temperatures used are the surface
temperature of the comet, $T_\mathrm{c}$, and for the Sun we use $T_\sun = 5772\,\mathrm{K}$ \citep{Hertel2015}. The geometric dilution factor for the Sun,
$\beta_\sun$, at 1 au is about $5\times 10^{-6}$ and at the surface of the comet we have that $\beta_\mathrm{c}(r_\mathrm{c}) = 1/2$.
We have not included thermal radiation from dust particles in the modelling presented here.

In order to solve the DEs, Eqs.~(\ref{eq:DE1}) and (\ref{eq:DE2}), we adopt initial population values, at $t=0$ and $r= r_\mathrm{c}$, appropriate to the
production rate, $n(0) = Q/(4\pi r_\mathrm{c}^2 v_\mathrm{e})$, and the initial excitation is entirely governed by the temperature of the comet surface,
$T_\mathrm{c}$. Following \citet{Bensch2004}, we adopt a constant neutral gas kinetic temperature in the coma cloud set to the $T_\mathrm{c}$ value.
We note that this is a simplification and some adiabatic cooling can take place \citep{Biver2015}.
We use a fourth-order Runge-Kutta scheme \citep{AS1972} when solving the DEs with adapting the step size because of the radial gradients.
The coma model cloud is assumed to be spherically symmetric. At any radius,
we can also determine level populations in the statistical equilibrium (SE) case similar to \cite{Paganini2010} by only including radiative and
collisional processes.

We are only modelling HCN and CH$_3$OH comet emission in this study. The overall photo-dissociation and photo-ionization rates for HCN, CH$_3$OH and H$_2$O (given at 1 au from the quiet Sun) used
come from the compilation by \citet{Huebner1992} which are essentially the same, for the molecules considered here, as those listed in the more recent work by \citet{Huebner2015}. Depending on the heliocentric distance $R_\mathrm{h}$ of the comet, these are scaled by $(1/ R_\mathrm{h})^2$. Furthermore, we assume $R_\mathrm{h}$ to be constant during the solution.

To properly account for collisional excitation we include collisions by electrons and H$_2$O molecules.
While the molecule-electron collisions are relatively easy to determine,
collisions by water molecules are much more difficult to compute. For instance, it is only very recently that collisional rate coefficients for the
collision systems CO-H$_2$O ($J\le 10$) have been determined accurately \citep{Faure2020}. In the case of HCN (with hyperfine structure) and CH$_3$OH, no such calculations exist yet. Limited ($J < 8$) HCN-H$_2$O collisional rate coefficients have been determined by \citet{Dubernet2019} which are being used here together
with the HCN-He rates of \citet{Dumouchel2010} for higher $J$ ($J\ge 8$). The latter rates were scaled to approximately match the $J < 8$ HCN-H$_2$O rates before being concatenated to the \citet{Dubernet2019} rates. For CH$_3$OH we
use the CH$_3$OH-H$_2$ rates of \citet{Rabli2010} (for para H$_2$). We treat the methanol $A$- and $E$-species separately and do not include
excited torsional states. The neutral collisional rates used are those valid for a kinetic temperature, $T_\mathrm{k}$, of 50~K and then
simply scaled with $(T_\mathrm{k}/50\, \mathrm{K})^{1/2}$ for other temperatures. We use the
tabulated downward collisional rates and calculate the corresponding upward rates from detailed balance.
For HCN-e collisional rates we adopt
those determined by \citet{Faure2007} and for the electron excitation of CH$_3$OH we use a set of rates computed in the Born approximation.

Normally, to speed up the calculations, we only include energy levels below 150-250~K for HCN and CH$_3$OH. The exact cut-off depends on the temperature. However, we have the option to include higher energy levels for testing
if there are any truncation problems. Furthermore, also vibrationally excited levels ($\nu_2$) for HCN can be fully included. However, the situation with a large number of levels together with levels with small populations requires much smaller step sizes to maintain
the solution accuracy throughout the coma. This makes the computation very time consuming and a much faster way is to
include effective pumping rates, see \citet{Paganini2010, Bensch2004}, via the $\nu_1$, $\nu_2$, $2\nu_2$, and $\nu_3$ vibrational states. We here only include the
$P$- and $R$-branches for the ro-vibrational HCN transitions since they can cause redistribution of the ground state
level populations in the outer coma \citep[e.g.][]{BM1987}. The $Q$-branch ($\Delta J = 0$) transitions do not do that by themselves (when
neglecting rotational relaxations within the excited state). This way of incorporating effective pumping rates, adopting the radiation field
in Eq.~(\ref{eq:J}), is only
valid when a small fraction of the molecules is in the excited states. This is typically the case for HCN \citep{BM1984}. We
have not used any pumping rates for CH$_3$OH in our modelling here.

To fully quantify the collisional processes as a function of cometocentric radius, not only the collisional rates are needed but also the radial
distributions of water molecules and electrons. The water distribution is assumed to follow the Haser equation (Eq.~\ref{eq:Haser}) assuming
a water production rate $Q_\mathrm{H_2O}$. The water production rate can be connected to the molecular (HCN or CH$_3$OH) production rate
$Q_\mathrm{mol}$ via the abundance ratio $Q_\mathrm{mol}/Q_\mathrm{H_2O}$ normally referred to as the mixing ratio.
To quantify the electron density radial distribution, $n_\mathrm{e}(r)$, we adopt the approach by \citet{Bensch2004} who based their formulation
on the work by \cite{Biver1997} \citep[see also][]{Zakharov2007}. These authors express $n_\mathrm{e}(r)$ (and electron temperature) in terms of the water production rate and like
\citet{Paganini2010} we use $x_{n_\mathrm{e}} = 0.3$ as the density scaling factor. The rather complex radial behaviour of the electron density and temperature is visualized
by \citet{Bensch2004} in their Fig.~1.

The solution of the time-dependent radiative transfer is in practice accomplished by a Python code called \texttt{comrad.py}. This code is
adapted to read molecular data files used in the Leiden atomic and molecular database \citep{Schoier2005} slightly modified to also include
photo-destruction values in the Solar radiation field at 1~au \citep{Huebner1992} and a section of
IR pumping transitions (for HCN $P$- and $R$-branches between the ground state and the excited states $\nu_1$, $\nu_2$, $2\nu_2$, and $\nu_3$). The frequencies and $A$-coefficients of the IR pumping transitions have been taken from the HITRAN molecular database \citep{Gordon2017}. After the molecular level populations have been determined
as a function of cometocentric radius, we obtain the modelled spectra, assuming a constant expansion velocity and a turbulent velocity width, by
convolving the projected radial intensity distribution with a gaussian appropriate to the telescope beam size and the comet
distance ($\Delta$).
Although the HCN level populations are strongly affected in the outer coma by the IR pumping transitions, the
impact on the resulting line intensities of ground state rotational transitions is small since the major contribution comes from
regions of high density (this can be pronounced by smaller observing beams).

In Appendix~\ref{app:transient} we will investigate the effects of the time-dependent radiative transfer calculations by
comparing the results to those of steady-state SE calculations. We will use the time-dependent radiative transfer model described above in the analysis of our HCN and
CH$_3$OH data and the results of the modelling are described in the following section.

\subsection{Model results}
\label{sec:modelresults}

Our HCN modelling results are summarized in Table~\ref{tab:models}. Here also the properties used for each comet
(46P/Wirtanen only 20 December 2018) are listed. The adopted nucleus radius, $r_\mathrm{c}$, which serves as a starting point for the model calculation, is also included but has essentially no impact on the modelling results.
For 46P and Y1 we have obtained the expansion velocities from the line widths. This results in $0.55\,\mathrm{km\, s^{-1}}$
for both comets. \citet{Coulson2020} found a similar value of $\sim 0.6\,\mathrm{km\, s^{-1}}$ for 46P
(at $R_\mathrm{h}=1.06\,\mathrm{au}$).
We note that, for 45P, \citet{Lovell2017} report an outflow velocity of about $0.8\,\mathrm{km\, s^{-1}}$ from OH observations (at $R_\mathrm{h}=0.54\,\mathrm{au}$).
\cite{Biver2006} determined a relation of $v_\mathrm{e}$
and heliocentric distance (in au): $v_\mathrm{e} \sim 0.8/\sqrt{R_\mathrm{h}}\,\mathrm{km\, s^{-1}}$. This relation yields expansion velocities near $1.3\,\mathrm{km\, s^{-1}}$ for 2P/Encke and around $0.55\,\mathrm{km\, s^{-1}}$ for 2I/Borisov. The observed $v_\mathrm{e}$ around $0.55\,\mathrm{km\, s^{-1}}$ for 46P and Y1 lie below this relation. For 46P and Y1 the above relation predicts $0.76\,\mathrm{km\, s^{-1}}$ and $0.70\,\mathrm{km\, s^{-1}}$, respectively.
However, for the other comets we have used an expansion velocity according to this relation.

In the case of U1 and O1 we have not
found representative water production rates in the literature. Here we instead used the mean value of
$Q_\mathrm{HCN}/Q_\mathrm{H_2O} = 0.1\mathrm{\%}$ from \citet{BM2017} to determine an appropriate $Q_\mathrm{H_2O}$
for the modelling.
It should be noted that the derived $Q_\mathrm{HCN}$ does not depend strongly on the adopted
$Q_\mathrm{H_2O}$ (as verified below by the 46P results where a factor of two increase in $Q_\mathrm{H_2O}$ resulted in a 7\% increase in the required $Q_\mathrm{HCN}$).

When possible we also use published values for the gas temperature (which also is used for the neutral gas
kinetic temperature here and $T_\mathrm{e}$ out to $R_\mathrm{CS}$) as indicated in the table. The kinetic
temperature is clearly dependent on the heliocentric distance $R_\mathrm{h}$ and \citet{Biver_etal1997}
estimated the kinetic temperature of C/1995 O1 (Hale-Bopp) from CH$_3$OH (and CO) observations as a function of
$R_\mathrm{h}$ to be around 100~K at 1~au and approximately scaling as $R_\mathrm{h}^{-1}$ with heliocentric distance.
In the case of 46P/Wirtanen we use our CH$_3$OH observations to estimate the kinetic temperature, see below.
If a kinetic temperature estimate is lacking, we used $60\,\mathrm{K}$ for comets with $R_\mathrm{h} \geq 1.3\,\mathrm{au}$ and
$70\,\mathrm{K}$ for the remaining comets in the model calculations.
These values approximately reflect the temperature behaviour as a function of heliocentric distance found by
\citet{Biver1999} for C/1992 B2 (Hyakutake) or by \citet{Disanti2016} in the case of D/2012 S1 (ISON).
The temperature estimates used in Table~\ref{tab:models} sometimes stem from rotational temperatures of molecules like H$_2$O or CH$_3$OH. How well this
rotational temperature reflects the gas kinetic temperature of the major collision agent (H$_2$O) depends on the degree of
thermalization. Also, collisions by electrons may change the excitation of the molecular probe \citep{Xie1992}.

\begin{table*}
\caption{\label{tab:models} Properties of the comets and results of the HCN modelling}
\centering
   \begin{tabular}{lcccccccc}
   \hline\hline
Comet                                   & $R_\mathrm{h}$  &   $\Delta$  &  $r_\mathrm{c}$ &  $v_\mathrm{e}$ & $T_\mathrm{c}$ & $Q_\mathrm{H_2O}$ & $Q_\mathrm{HCN}$  & $Q_\mathrm{HCN}/Q_\mathrm{H_2O}$ \\
                                        & (au) &    (au)     &       (km)     & (km/s) &      (K)       & (mol/s) &    (mol/s)    &  (\%)  \\
\hline
C/2016 U1 (NEOWISE)                     & 0.50 &  0.90  & 1.0 & 1.1 & 70 & ... &  $< 7.8\times 10^{25}$ & ... \\
45P/Honda-Mrkos-Pajdu\u{s}\'{a}kov\'{a} & 0.88 &  0.12  & 0.4\tablefoottext{a} & 0.8 & 70 & $2\times 10^{27}$\tablefoottext{b} & $< 3.2\times 10^{24}$  & $<0.16$ \\
2P/Encke                                & 0.38 &  0.70  & 2.2\tablefoottext{c} & 1.3 & 70\tablefoottext{d} & $3\times 10^{28}$\tablefoottext{d} & $< 2.2\times 10^{26}$ & $< 0.7$ \\
41P/Tuttle-Giacobini-Kres\'{a}k         & 1.05 &  0.15  & 0.6\tablefoottext{e} & 0.8 & 70 & $3.5\times 10^{27}$\tablefoottext{f} & $(4.5\pm 1.0)\times 10^{24}$ & $0.13\pm 0.03$ \\
C/2015 ER61 (PanSTARRS)                 & 1.10 &  1.22  & 0.94\tablefoottext{c} & 0.7 & 70\tablefoottext{g} & $1\times 10^{29}$\tablefoottext{g} & $(8.2\pm 2.1)\times 10^{25}$ & $0.082\pm 0.021$ \\
C/2017 E4 (Lovejoy)                     & 0.63 &  0.67  & 1.0 & 1.0 & 70 & $3\times 10^{28}$\tablefoottext{h} & $< 6.3\times 10^{25}$ & $<0.21$ \\
C/2015 V2 (Johnson)                     & 1.67 &  0.87  & 1.7\tablefoottext{c} & 0.6 & 60 & $4\times 10^{28}$\tablefoottext{i} & $(3.5\pm 0.9)\times 10^{25}$ & $0.088\pm 0.022$ \\
46P/Wirtanen (2018-12-20)               & 1.06 &  0.08  & 0.7\tablefoottext{j} & 0.55 & 70\tablefoottext{k} & $1.6\times 10^{28}$\tablefoottext{j} & $(1.6\pm 0.1)\times 10^{25}$ & $0.10\pm 0.01$ \\
                                        &      &        &                    &      &               & $8\times 10^{27}$\tablefoottext{l} & $(1.5\pm 0.1)\times 10^{25}$ & $0.19\pm 0.02$ \\
C/2017 O1 (ASASSN1)                     & 1.50 &  0.72  & 0.92\tablefoottext{c} & 0.6 & 60 & ... & $(2.2\pm 0.2)\times 10^{25}$ & ... \\
C/2018 Y1 (Iwamoto)                     & 1.30 &  0.42  & 1.0 & 0.55 & 60 & $2.1\times 10^{28}$\tablefoottext{m}  & $(2.2\pm 0.2)\times 10^{25}$ & $0.10\pm 0.02$ \\
2I/Borisov                              & 2.10 &  2.40  & 1.0 & 0.55 & 50\tablefoottext{n} & $6.5\times 10^{26}$\tablefoottext{n} & $< 6.3\times 10^{24}$ & $<1.0$ \\
 \hline
    \end{tabular}
\tablefoot{
\tablefoottext{a}{From \cite{Combi2019}.}
\tablefoottext{b}{From \citet{DelloRusso2020}.}
\tablefoottext{c}{Comet radius from \citet{Paradowski2020}.}
\tablefoottext{d}{From \citet{Roth2018}.}
\tablefoottext{e}{See \cite{Howell2017} and \cite{Boehnhardt2020}.}
\tablefoottext{f}{April 2017 average from \citet{Moulane2018}.}
\tablefoottext{g}{Adapted from \citet{Saki2021}.}
\tablefoottext{h}{From \citet{Faggi2018}.}
\tablefoottext{i}{From \cite{Combi2021}.}
\tablefoottext{j}{Adapted from \cite{Combi2020}.}
\tablefoottext{k}{Estimated from the CH$_3$OH data in this paper.}
\tablefoottext{l}{Based on IR spectroscopy data \citep{Bonev2021}. See also \citet{Lis2019}.}
\tablefoottext{m}{Preliminary data by \citet{Disanti2019}.}
\tablefoottext{n}{Adapted from \citet{Cordiner2020}.}
}
\end{table*}

Due to the observed blue-shifted line profiles for 46P/Wirtanen and also C/2018 Y1 (Iwamoto) it is likely that there is an anisotropic distribution of $Q_\mathrm{HCN}$
in the sense that the production rate is higher on comet sunward side than on the anti-sunward side (with a factor of 2-3 difference, see Fig.~\ref{fig:wirtanen}). This was also noted by \citet{Wang2020} and \citet{Biver2021} for HCN and for CH$_3$OH by \citet{Roth2021b}. Our modelled values, based on a spherically symmetric model geometry, thus refer to an average production rate.

As already mentioned, in the case of 46P/Wirtanen we can estimate the gas kinetic temperature since we have detected lines with quite different
upper energies (as high as 83 K). Another aspect of CH$_3$OH transitions, due to the nuclear spin directions of the methyl
group H atoms, is that they come in two different symmetry species, $A$ and $E$, which are, as pointed out by \citet{BM1994},
radiatively and collisionally uncoupled apart from possible line overlaps. The lowest $E$-state is about 8~K above
the lowest $A$-state. Depending on the formation mechanism they may be produced in unequal amounts.
This can be the case at low temperatures \citep[e.g.][]{Wirstrom2011}. If the temperature appropriate for
methanol formation in comets is larger than 8~K,
then we expect the production rates of the $A$- and $E$-species of CH$_3$OH to be equal. We have performed a grid testing
with temperatures ranging from 40~K to 120~K (in steps of 10~K) and $Q_\mathrm{CH_3OH}$ from $1.0\times 10^{26}\,\mathrm{mol\, s^{-1}}$ to
$3.0\times 10^{26}\,\mathrm{mol\, s^{-1}}$ (in steps of $0.1\times 10^{26}\,\mathrm{mol\, s^{-1}}$). The water production rate used was
$Q_\mathrm{H_2O}=1.0\times 10^{28}\,\mathrm{mol\, s^{-1}}$ which is an average of the $Q_\mathrm{H_2O}$ determined by
\citep{Combi2020} for the dates 22-28 December 2018.
The grid combination of kinetic
temperature and $Q_\mathrm{H_2O}$ that gave
the best fit (as indicated by smallest $\chi^2$-value) to the observed integrated intensities in Table~\ref{tab:ch3oh_results} was for a temperature of $70\pm 15\,\mathrm{K}$ and
$Q_\mathrm{CH_3OH} = (1.6\pm 0.1)\times 10^{26}\,\mathrm{mol\, s^{-1}}$ or 1.6\% of the water production rate.

The fitting indicated, within the errors, also that the CH$_3$OH $A$- and $E$-species are being outgassed in equal amounts in 46P/Wirtanen.
The resulting model spectra are shown in Fig.~\ref{fig:wirt_ch3oh} together with the observed spectra.

\begin{figure*}
\sidecaption
\includegraphics[width=12cm]{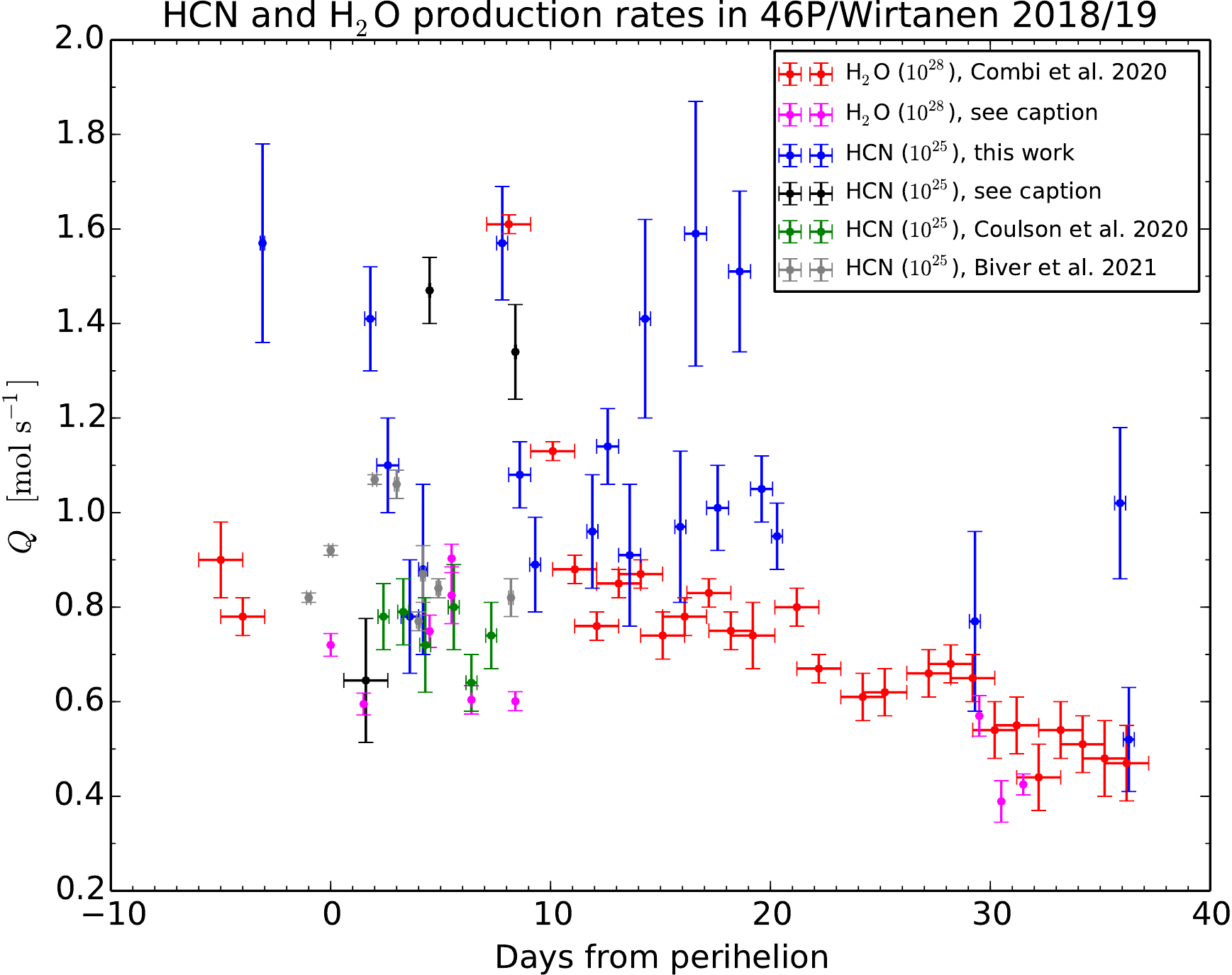}
\caption{HCN and H$_2$O production rates in 46P/Wirtanen as a function of days from perihelion (12.9 December 2018 UT). The HCN rates (blue) have been determined from the
intensities (only those detected above the 3$\sigma$ level) in Table~\ref{tab:wirtanen_results}. The water production rates (red) contemporary with our $Q_\mathrm{HCN}$ results, are from \citet{Combi2020}. The perihelion value of $Q_\mathrm{H_2O}$ from \citet{Moulane2019} and the post-perihelion values from \citet{Saki2020} (14 and 19 December), \citet{Roth2021a} (18 December), \cite{Bonev2021} (17/18 December), \citet{Khan2021} (21 December), and \cite{McKay2021} (11--13 January ) are included (magenta). Also, with black markers, the $Q_\mathrm{HCN}$ as estimated by \citet{Wang2020} (14/15 December), \cite{Bonev2021} (17 December) and \citet{Khan2021} (21 December) are shown.  The HCN observations by \cite{Coulson2020} have been included using green markers and those from \citet{Biver2021} are shown with grey markers. The lengths of the error bars in
the vertical direction relate to the 1$\sigma$ uncertainties in the rates while the in horizontal direction they
reflect the time length over which they refer to. As indicated, the HCN rates should be scaled by $10^{25}$ and
the water rates by $10^{28}$.
}
\label{fig:QHCN_QH2O}
\end{figure*}

As noted above, the CH$_3$OH modelling of 46P/Wirtanen resulted in a kinetic temperature of 70 K. Using this temperature and
$Q_\mathrm{H_2O}=1.6\times 10^{28}\,\mathrm{mol\, s^{-1}}$ which is the 22 December value of \citet{Combi2020}, we find that for the 20 December data an
HCN production rate of $(1.6\pm 0.1)\times 10^{25}\,\mathrm{mol\, s^{-1}}$ is matching the observed integrated intensity,
see Tables~\ref{tab:models} and \ref{tab:wirtanen_results}. Using a fixed value for the ratio $Q_\mathrm{HCN}/Q_\mathrm{H_2O}$ of
0.1\%, as found for the 20 December data, we have run models for 3 different HCN production rates at different dates and
entered the expected 1-0 integrated intensity in Fig.~\ref{fig:wirt_hcn}. The production rates are
$(0.5, 1.0, 1.5)\times 10^{25}\,\mathrm{mol\, s^{-1}}$ which encompass most of the observed line intensities. Also included in Table~\ref{tab:models}
is the HCN production rate when adopting a lower water production rate as indicated by the infrared spectroscopy results \citep[e.g.][]{Bonev2021} a few days prior to
20 December. The change in $Q_\mathrm{HCN}$ is less than 10\%.
The individual HCN production rates, see Table~\ref{tab:wirtanen_results}, are shown in Fig.~\ref{fig:QHCN_QH2O} together with the water production rates
obtained using SOHO/SWAN by \citet{Combi2020}. These results will be discussed in the next section.

Our CH$_3$OH observations of 96P/Machholz only resulted in an upper limit, see Table~\ref{tab:ch3oh_results}. Adopting
$T_\mathrm{c}=90\,\mathrm{K}$ and $Q_\mathrm{CH_3OH}/Q_\mathrm{H_2O} = 1\%$ this upper limit corresponds to a very high methanol
production rate. We estimate the upper limit production rate to be above $10^{28}\,\mathrm{mol\, s^{-1}}$. The poor constraining of $Q_\mathrm{CH_3OH}$ is due
to the fact that 96P was only at $R_\mathrm{h} = 0.12\,\mathrm{au}$ at the time of the measurements and the size of the neutral coma is
considerably smaller due to the effective destruction of H$_2$O and CH$_3$OH molecules near the Sun.

\section{Discussion}
\label{sect:dis}

We will in this section first discuss the previously determined HCN production rates as well as the upper limits. We will also discuss the HCN production rates as an indirect indicator of water production rates via mixing ratios. In the case of 46P/Wirtanen, the monitoring results can be used to see variations in the HCN production rates on time scales of 1 day and these will be compared to other observations of HCN and H$_2$O production rates. Our usage of methanol as a thermal probe will also be discussed. Finally, our time-dependent treatment of the radiative transfer of comets will be discussed in relation to a more standard approach performed in the steady-state limit of the SE assumption.

\subsection{HCN production rates}
As can be seen in Table~\ref{tab:models} we have detected HCN in 2 JFCs (41P and 46P, the latter comet is discussed in more detail in the next section) and 4 long-period OCCs (ER61, V2, O1, and Y1).
For the 6 comets with HCN detections, the determined HCN
production rates fall in the range $(0.5-8)\times 10^{25}\,\mathrm{mol\, s^{-1}}$ and there is no obvious difference
between the two type of comets. For the 5 comets where we also have independent (and reasonably contemporary) water production rates, the
HCN to H$_2$O mixing ratios are in the range 0.08-0.13\%. This range of
mixing ratios is consistent with the values as determined from radio observations and discussed by \citet{MummaCharnley2011} and by \cite{BM2017}. However it is smaller, by a factor of
about 2, than the typical value in \citet{DelloRusso2016}. The latter study, based on high-resolution infrared spectroscopy, compiles a typical range
of mixing ratios of 0.15-0.27\%. Our ratios, near 0.1\% for 41P, ER61, V2, 46P, and Y1, are
outside this range but are well above the lower extreme value of 0.03\% found by \citet{DelloRusso2016}.

In the case of 41P our HCN
data were obtained over a longer period in April 2017. The average water production rate, derived from OH observations by
 \citet{Moulane2018} over this period, is estimated to be about $3.5\times 10^{27}\,\mathrm{mol\, s^{-1}}$ with no larger variations.
 \citet{Moulane2018} also report CN production rates in the range $(4-5)\times 10^{24}\,\mathrm{mol\, s^{-1}}$ which is similar
 to our HCN production rate of $(4.5\pm 1.0)\times 10^{24}\,\mathrm{mol\, s^{-1}}$. This could indicate that
 the CN radicals observed in 41P mainly stem from photo-dissociation of HCN.
 However, given the uncertainties and that our $Q_\mathrm{HCN}$ is an average (over time and over the coma) one cannot exclude a contribution to $Q_\mathrm{CN}$ from
 dust grains as suggested by \citet{DelloRusso2016} and \citet{Hanni2020}.

The ER61 observations (HCN and H$_2$O) by \citet{Saki2021} occurred on three occasions in our 6-22 April observation range and we here
adopt an average of their water production rates of $1\times 10^{29}\,\mathrm{mol\, s^{-1}}$. They deduced HCN mixing ratios in the range 0.11-0.14\%
somewhat higher than our value of 0.082\%. This is possibly a result of that our HCN data reflect a larger time span.
However, the ALMA results \citep{Roth2021c} indicate an HCN mixing ratio (0.072\%) consistent with our value.

For Y1, the used water production rate of \citet{Disanti2019} dates from about a week before our estimate of the HCN production rate. Adopting this $Q_\mathrm{H_2O}$, the mixing ratio is $0.10\pm 0.02\,\%$. \citet{Disanti2019} obtain an HCN mixing ratio of 0.2\% on 4 February 2018. This could indicate variations in $Q_\mathrm{H_2O}$ if we assume that
the HCN production rate is constant over this period of observations.

In the case of the remaining HCN detections, V2 and O1, no other directly determined molecular production rates have been found to date in the literature.
However, for V2, \cite{Combi2021} estimated the water production rate, using SOHO/SWAN observations, to about $4\times 10^{28}\,\mathrm{mol\, s^{-1}}$ near our observation date. The corresponding mixing ratio is about 0.08\%.
We also note that \citet{Venkataramani2018} reported on the absence of molecular emissions towards V2 at $R_\mathrm{h}=2.83\,\mathrm{au}$ but they noted that such emissions appeared
later on a low level when $R_\mathrm{h}=2.3\,\mathrm{au}$. Our $3.9\sigma$ detection was made at $R_\mathrm{h}=1.67\,\mathrm{au}$.

In 5 of our observed comets (U1, 45P, 2P/Encke, E4, and 2I/Borisov), we only obtained upper limits on $Q_\mathrm{HCN}$, see Table~\ref{tab:models}.
At the time of writing, no molecular observational results for U1 have been reported in the literature so here we have nothing to compare to.
The HCN production rate of $(3-4)\times 10^{24}\,\mathrm{mol\, s^{-1}}$ in 45P found by
\citet{DelloRusso2020} refers to dates in mid February 2017. Our non-detection, $Q_\mathrm{HCN} < 3.2\times 10^{24}\,\mathrm{mol\, s^{-1}}$, is from early February and indicates that the production rate of HCN was about same or lower in early February.
\citet{Roth2018} observed HCN in 2P/Encke, after perihelion, on 21 and 25 March, resulting in $Q_\mathrm{HCN} \sim (3-6)\times 10^{25}\,\mathrm{mol\, s^{-1}}$. Our upper limit only indicates that $Q_\mathrm{HCN}$, before and around perihelion, did not exceed 6-8 times this value.
For E4, \cite{Faggi2018} determined an HCN production rate of $5\times 10^{25}\,\mathrm{mol\, s^{-1}}$. This was 3 days before we
obtained an upper limit of $6\times 10^{25}\,\mathrm{mol\, s^{-1}}$ so apparently there was no significant increase in $Q_\mathrm{HCN}$ from April 4 to 7.
Our upper limit on $Q_\mathrm{HCN}$ towards 2I/Borisov is consistent with the ALMA HCN detection on 14/15 December
by \citet{Cordiner2020}. They determined that
$Q_\mathrm{HCN} = 7\times 10^{23}\,\mathrm{mol\, s^{-1}}$.

Taken together, the HCN mixing ratios, based on our $J=1-0$ data, seem to fall near 0.1\% when we have reasonably contemporary estimates of $Q_\mathrm{H_2O}$.

\subsection{Variation in the HCN production rates for 46P/Wirtanen}
We determined HCN production rates for 46P/Wirtanen (Table~\ref{tab:wirtanen_results}) from a few days before perihelion until 36 days after perihelion, see
Figs.~\ref{fig:wirt_hcn} and \ref{fig:QHCN_QH2O}. There are clear variations in $Q_\mathrm{HCN}$, sometimes on a daily basis, but it is also evident that $Q_\mathrm{HCN}$
has decreased by a factor of 2-3 some month after the perihelion passage. We note that \citet{Wang2020} observed HCN 1-0 on 14-15 December and obtained a $Q_\mathrm{HCN}$ (based on LTE calculations) of about half our value and that the IRAM 30-m observations \citep{Biver2021} on 14 December result in a $Q_\mathrm{HCN}$ in between our value ($1.4\times 10^{25}\,\mathrm{mol\, s^{-1}}$) and that of \cite{Wang2020}. A few days
later, in the range 15-17 December, the IRAM 30-m data show $Q_\mathrm{HCN}$ determinations very close to our values.
The James Clerk Maxwell Telescope (JCMT) HCN(4-3) data of \citet{Coulson2020} showed little daily variation in
$Q_\mathrm{HCN}$ in the period 15 December to the early UT hours of 20 December.  The HCN production rates determined by \citet{Wang2020}, \citet{Biver2021} and \citet{Coulson2020} are all included in Fig.~\ref{fig:QHCN_QH2O}. The HCN(4-3) $Q_\mathrm{HCN}$ results agree quite well with
our results up until 20 December. Here we see a clear increase in $Q_\mathrm{HCN}$ to $(1.6\pm 0.1)\times 10^{25}\,\mathrm{mol\, s^{-1}}$, by a factor of two over that from
\citet{Coulson2020}. However, the JCMT observations relate to the early UT hours and our observations are from about 12 hrs later the same day. This would indicate an outburst event during the day of 20 December. However, early on 21 December the zero-spacing NOEMA observations of \citet{Biver2021} again indicates a lower $Q_\mathrm{HCN}$ of $0.8\times 10^{25}\,\mathrm{mol\, s^{-1}}$ so if there was an outburst it must have been rather short ($\la 12\,\mathrm{hr}$) not to be recorded by the NOEMA observations.
In fact, similar changes of $Q_{\mathrm{CH_3OH}}$, by about a factor 2, was seen by \citet{Roth2021b} connecting it to the rotation period time scale of 9 hrs.

Based on infrared spectroscopy, \citet{Bonev2021} reported an HCN production rate of $(1.47\pm 0.07)\times 10^{25}\,\mathrm{mol\, s^{-1}}$ on 17 December. This $Q_\mathrm{HCN}$ is
significantly higher than our value of $(0.9\pm 0.2)\times 10^{25}\,\mathrm{mol\, s^{-1}}$ and the value, $(0.7\pm 0.1)\times 10^{25}\,\mathrm{mol\, s^{-1}}$, of \citet{Coulson2020} from the same date. Four days later, on 21 December, \citet{Khan2021} observed HCN also by infrared spectroscopy. Their HCN production rate averaged to about
$1.3\times 10^{25}\,\mathrm{mol\, s^{-1}}$ which is near our value, see Table~\ref{tab:wirtanen_results}, $1.1\times 10^{25}\,\mathrm{mol\, s^{-1}}$ on the same date. The $Q_\mathrm{HCN}$ values from \citet{Bonev2021} and \citet{Khan2021} have been included in Fig.~\ref{fig:QHCN_QH2O}.

As already pointed out in Sect.~\ref{sec:modelresults}, there appears to be a clear difference, a factor of 2-3, in the outgassing activity on the sunward vs anti-sunward parts of the comet nucleus over the course of about one month. As reported by
\citet{Biver2015} based on their water line observations by the Microwave Instrument for the Rosetta Orbiter (MIRO) of the JFC 67P/Churyumov-Gerasimenko, the nightside water production rate was low, $< 1\%$ of the dayside production rate (at $r_\mathrm{h}=3.4\,\mathrm{au}$). Later, at $r_\mathrm{h}=1.8\,\mathrm{au}$, \citet{Fink2016} report for 67P that about 17\% of
the total water production emerges from the nightside. In any case, this is a higher
day-to-nightside activity ratio than we see for 46P but our ratio is averaged over a period of a month in which the daytime fraction as
seen by our telescope beam changed (elongation varied from 160 to $140\deg$ from 20 December 2018 to 18 January 2019).

The SOHO/SWAN monitoring observations \citep{Combi2020} to estimate $Q_\mathrm{H_2O}$ in 46P coincide partly in time with our
observations and provide estimates of the water production rate variation. Since these results are based on Ly-$\alpha$ observations of the hydrogen coma (dissociation product from H$_2$O and OH), they refer to an average water production of the 2-3 days prior to the actual observation date (as indicated by the horizontal error bars in Fig.~\ref{fig:QHCN_QH2O}) so the $Q_\mathrm{H_2O}$ determined this way may not probe activity variations on shorter time scales. However, the trend we see in $Q_\mathrm{HCN}$ from 20 December until 18 January (8 to 36 days after perihelion) is consistent with a mixing ratio near 0.1\%. For the week preceding this time span, \citet{Biver2021} report essentially the same value $0.11\pm 0.01\,\%$. There are a few $Q_\mathrm{HCN}$ estimates in our monitoring time span 8-36 days after perihelion that seem to indicate an HCN mixing ratio near 0.2\%. This is also the case for the pre-perihelion observations. To distinguish whether this reflects real changes in the mixing ratio, by a factor of about 2, or is an effect of different time-averaging would need $Q_\mathrm{H_2O}$ measurements with $\sim$1 day time-resolution or better. If it
is an effect of time-averaging, the HCN production rate will be a good indicator of the water production rate
in the case of 46P/Wirtanen. The increase in $Q_\mathrm{HCN}$ seen by us on 20 December seems supported by the large water
production rate reported by \citet{Combi2020} 2-3 days later and, as discussed earlier, the $Q_\mathrm{HCN}$ reported by \citet{Khan2021} on 21 December but not by the NOEMA
observations early UT hours on 21 December.

\subsection{Coma gas kinetic temperature and methanol production in 46P/Wirtanen}
In Sect.~\ref{sec:modelresults}, we determined the gas kinetic temperature of 46P, from methanol data taken over the time span 22-28 December, to be $70\pm 15\,\mathrm{K}$. Furthermore, the deduced production rate of $Q_\mathrm{CH_3OH} = (1.6\pm 0.1)\times 10^{26}\,\mathrm{mol\, s^{-1}}$ corresponds to a methanol mixing ratio of
about 1.6\%. Our mixing ratio is very near the JFC average value of 1.7\% as compiled by \citet{DelloRusso2016} and
slightly less than the JFC median value of about 2\% as compiled by \citet{MummaCharnley2011}.

Most other methanol observations of 46P refer to the period around or the week after the perihelion. For instance, on 7-9 December \cite{Roth2021b} used ALMA observations
to determine methanol rotation temperatures in the range from 50~K to 80~K. \citet{Biver2021} determined, using IRAM 30-m CH$_3$OH observations, 6-day
averages (over the range 12-18 December) of the gas temperature in the range 53-75 K.
On 16 December \citet{Coulson2020} used the JCMT to observe three methanol lines around 338~GHz. They determined the rotation temperature to
$T_\mathrm{rot} \sim 30-50\,\mathrm{K}$.
Using infrared spectroscopy observations on 18 December, \cite{Roth2021a} report a methanol rotation temperature of about 90~K. In addition, \citet{Biver2021} also made a
single methanol observation on 25.8 December 2018 UT, within our observing time range, which resulted in a gas temperature of $43\pm 7\,\mathrm{K}$.

Three studies estimate H$_2$O rotation temperatures of 46P. On 14 and 19 December, \cite{Saki2020} determined the water rotation temperature to be about 85~K. On 18 December,
\cite{Roth2021a} report a rotation temperature of about 90~K. Later, on 21 December \citet{Khan2021} derived temperatures of 80~K to 90~K.
Our derived value of the gas kinetic temperature of about 70~K, as averaged over 22-28 December, is consistent with the studies mentioned above. There is a possibility that
the temperature has decreased from 80--90~K to about 40-50~K during our observations \citep[cf.][]{Khan2021, Biver2021}. However, this notion is somewhat complicated by the
variation of temperature estimates (30--90~K) prior to our observations.

In the case of the methanol production in 46P, we begin with noting that on 7-9 December, the ALMA CH$_3$OH observations \citep{Roth2021b} resulted in varying CH$_3$OH production
rates in the range $(2.0-3.6)\times 10^{26}\,\mathrm{mol\, s^{-1}}$.
The IRAM 30-m data by \citet{Biver2021} determined $Q_\mathrm{CH_3OH}$, averaged over 12-18 December, to $(2.6\pm 0.2)\times 10^{26}\,\mathrm{mol\, s^{-1}}$. The
JCMT observations on 16 December \citep{Coulson2020} resulted in $(3.5\pm 0.2)\times 10^{26}\,\mathrm{mol\, s^{-1}}$. On 18 December, \cite{Roth2021a} report
a methanol production rate around $2\times 10^{26}\,\mathrm{mol\, s^{-1}}$. A few days later, on 21 December, \citet{Khan2021} reported a similar methanol production
rate of $2.5\times 10^{26}\,\mathrm{mol\, s^{-1}}$. These estimates of $Q_\mathrm{CH_3OH}$ are all higher than our later, 22--28 December, value of $Q_\mathrm{CH_3OH} = (1.6\pm 0.1)\times 10^{26}\,\mathrm{mol\, s^{-1}}$. The NOEMA observations \citep{Biver2021} on 25.8 December 2018 UT resulted in $(1.84\pm 0.14)\times 10^{26}\,\mathrm{mol\, s^{-1}}$ close to our value.

As noted above, we adopt $Q_\mathrm{H_2O}=1.0\times 10^{28}\,\mathrm{mol\, s^{-1}}$ which is an average of the $Q_\mathrm{H_2O}$ determined by
\citep{Combi2020} for the dates 22-28 December 2018. This $Q_\mathrm{H_2O}$ results in a mixing ratio of $1.6\%$. For the 2 week period prior to our methanol observations, mixing ratios
in the range 2--5\% have been reported \citep{Roth2021b, Biver2021, Coulson2020, Roth2021a, Khan2021}.
Based on this, we suggest that the average methanol mixing ratio
decreased by about a factor two from around perihelion into our observing period of 22-28 December.

The production rates
of the methanol $A$- and $E$-species are about equal for 46P within the uncertainties. This would point at $\ga 8\,\mathrm{K}$
temperature environment during the formation of the methanol molecules. Finally, the determined kinetic temperature is
related to the used CH$_3$OH-H$_2$O collisional rate coefficients which in our case were based on the CH$_3$OH-H$_2$ rates.
As already pointed out, accurate collisional rate coefficients currently only exist for the collision systems
CO-H$_2$O \citep{Faure2020} and HCN-H$_2$O \citep{Dubernet2019}.

\subsection{Time-dependent radiative transfer versus SE calculations}
To study time-dependent effects, we used fragment B of 73P as test case (see Appendix~\ref{app:transient}) and we noted that
HCN level population deviations relative steady-state SE-calculations were of the order 5-15\% for $J=0$ to 5. The water production rate for 73P/B is around
      $2\times 10^{28}\,\mathrm{mol\, s^{-1}}$. We also found that the time-dependent deviations would be relatively larger for
comets with lower water production rates since the collisions with water molecules will be less frequent in the inner part of
the coma. Likewise, by reducing the effect of electron-HCN collisions the time-dependent effects became more pronounced. For molecules with lower electric dipole moments (and hence generally lower $A$-coefficients), we expect the effects to be more pronounced - at least in the outer
coma where radiative processes dominate (see Fig.~\ref{fig:73P_hcn}). However, strong IR-pumping could counteract the effect in
the outer coma making the details of the excitation picture very complicated. We expect that the influence of time-dependent effects on the
 HCN level populations will be small ($\la$20\%) for $Q_\mathrm{H_2O} \ga 10^{28}\,\mathrm{mol\, s^{-1}}$.

It is worth pointing out that the total molecular population is bound to follow the radial Haser distribution so that molecular production rates based on
the analysis of a larger number of transitions should be less affected by the time-dependent effects encountered here.

Also, the present modelling work neglects
excitation effects at a certain position in the coma caused by line radiation from other parts of the coma (same as
assuming optically thin radiation in the level population determination). This may be a good approximation for most cometary
molecular probes but an important exception is water excitation \citep[e.g.][]{BM1987, Bensch2004}. Extending our model into
using an escape probability formalism \citep[cf.][]{Gersch2014} is a possibility but would need an iterative solution technique. That is; solving
for the entire coma using an initial guess (using the DE here or simply an SE solution), include the line radiation in Eq.~(\ref{eq:J}), and then repeat the procedure with time-dependent solutions throughout the coma until the level populations converge. Presumably, this
would be a very time consuming technique but could be tried for water excitation in comets. Solution methods based
on the Monte Carlo technique \citep{Bernes1979} or those using accelerated lambda iteration \citep{Rybicki1991, Bergman2020} appear to be less suitable for time-dependent studies since they rely on the SE steady state assumption. However, in the escape probability formalism \citep[e.g.][]{BM1987, vanderTak2007, Zakharov2007}, the
$A$-coefficients are replaced by $\beta A$ where $\beta$ is the (local) escape probability for the transition in question.
In the case of optically thick transitions ($\beta \sim 1/\tau$) this could imply that time-dependent effects, due a reduction in
radiative excitation, could be of a relatively higher importance.

Another aspect is that because the radiative transfer formulation here is in the time domain, it would be easier to incorporate chemical and photolytic state-to-state reactions
when solving excitation of several molecules at the same time in the coma. Finally, it is worth stressing that most radiative transfer models,
including the one presented here, assume a spherically symmetric coma geometry. Remote observations and in-situ studies of cometary comae clearly
show deviations from this geometry and models like that of \citet{Debout2016} could be used for studying time-dependent effects also in anisotropic comae.

\section{Summary and conclusions}
\label{sect:con}
We performed spectral line observations of HCN and CH$_3$OH transitions towards a sample of bright comets using the OSO 20-m telescope. From the
observations, we determined molecular production rates using a radiative transfer model taking into account time-dependent effects. Our main conclusions are:

\begin{itemize}

\item We detected HCN(1-0) in 6 comets, 2 JFCs and 4 OCCs, of our sample. For 5 of these comets we could determine the HCN mixing ratio using published values of $Q_\mathrm{H_2O}$. The determined HCN mixing ratios were all near 0.1\% (0.08-0.013\%).

\item We detected HCN emission in 46P/Wirtanen on 21 occasions from 9 December 2018 to 18 January 2019. When comparing
with contemporary $Q_\mathrm{H_2O}$ determinations we find a typical mixing ratio near 0.1\% with possible variations up to 0.2\%. This
variation can also be due to short-time activity variations ($\la 1\,\mathrm{d}$) not reflected in the Ly-$\alpha$ observations used for determining $Q_\mathrm{H_2O}$. We also noted a clear asymmetry in the observed HCN and CH$_3$OH line profiles in the sense that the sunward activity was about a factor of 2-3 as strong as the outgassing activity on the 46P night-side.

\item In addition to HCN, we detected 6 CH$_3$OH transitions in 46P/Wirtanen around two weeks after perihelion in December 2018.
This enabled us to determine the gas
kinetic temperature to be about $70\,\mathrm{K}$. The methanol mixing ratio was found to be $1.6\%$ about half that of the methanol mixing ratio during the 2 weeks prior to our observations.

\item To interpret our HCN and CH$_3$OH observations, we used a radiative transfer code in which we allowed for time-dependent
effects on the level populations by relaxing the normal steady-state assumption of statistical equilbrium (SE) in the cometary coma. In our test case (73P fragment B, see Fig.~\ref{fig:73P_rpop})
we noticed 5-15\% deviations of the HCN level populations as compared to SE values. We recognized that the time-dependent effects
may be more pronounced for comets with lower water production rates and if the molecular excitation via electron collisions becomes
less efficient. For $Q_\mathrm{H_2O} \ga 10^{28}\,\mathrm{mol\, s^{-1}}$ the time-dependent effects in the HCN excitation should be small.

\end{itemize}

\begin{acknowledgements}
We are very grateful to L. Paganini for sharing his HCN modelling details. We also thank N. Biver for making his
1997 thesis available to us. The Onsala Space Observatory national research infrastructure is funded through Swedish Research Council grant No 2017-00648.
\end{acknowledgements}


\bibliography{aa_comet} 
\bibliographystyle{aa} 

\begin{appendix}
\section{Validation of the time-dependent model}
\label{app:transient}
\begin{figure}
\resizebox{\hsize}{!}{
\includegraphics{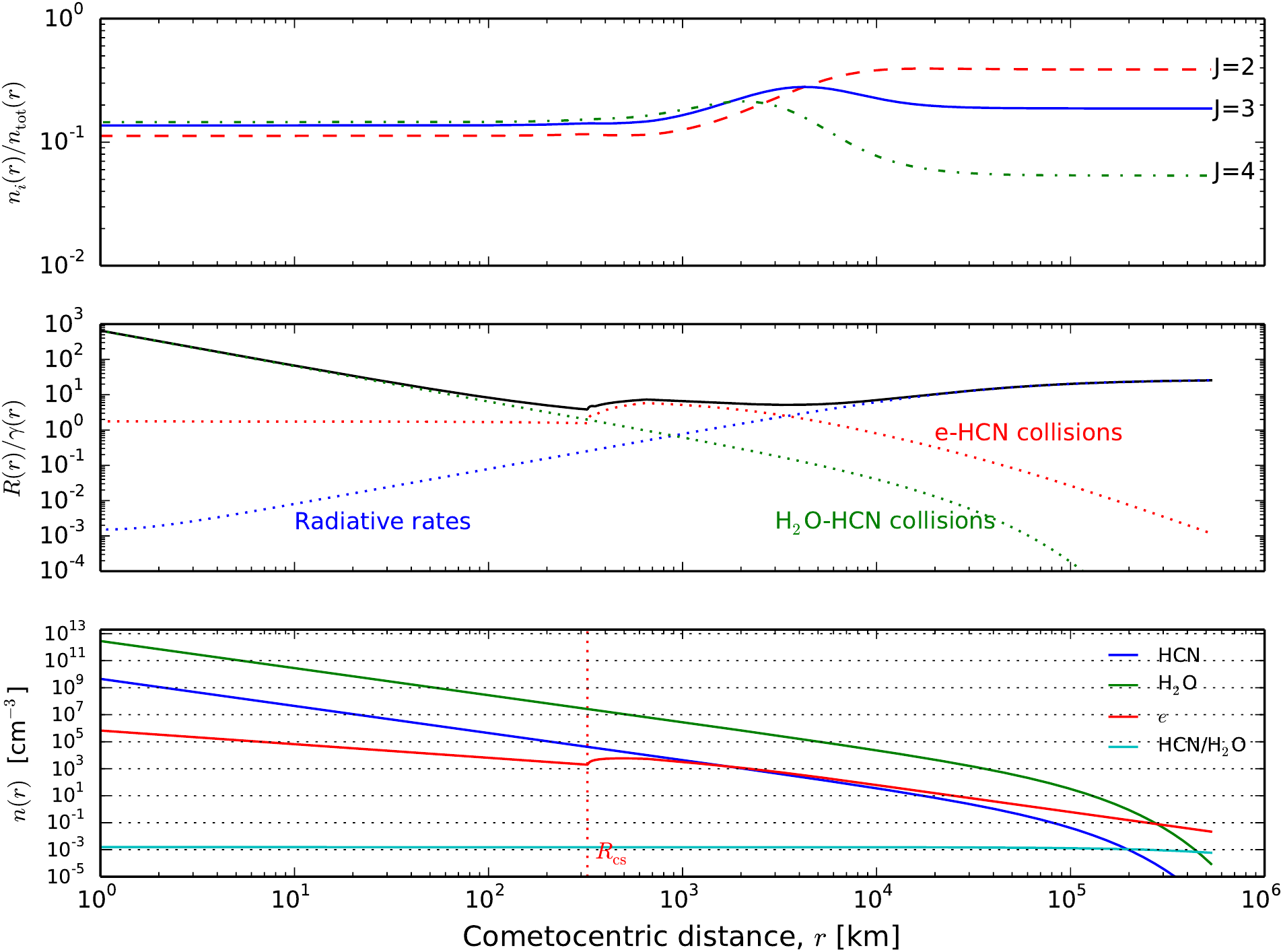}
}
\caption{Time dependent HCN modelling with parameters corresponding to the fragment B of 73P/Schwassmann-Wachmann.
The upper plot shows the fractional level populations for levels $J=2, 3, 4$ as a function of cometocentric distance.
The middle plot shows the relative importance of different downward rates for the HCN $J=3-2$
transition. As
indicated, the included rates are radiative (excluding IR pumping rates), and collisional rates both for electrons and
water molecules. The sum of these rates is included as a black line. They are normalized with the time dependent rate
due to destruction and constant expansion, see Eq.~(\ref{eq:gamma}).
The bottom panel depicts the radial density variations.
The blue curve is the total HCN density and the green curve represents the water density behaviour. The electron density,
$n_\mathrm{e}(r)$, is drawn in red. Also the HCN/H$_2$O ratio is shown. It is not entirely flat since when the photo-dissociation
becomes important in the outer coma, HCN and H$_2$O molecules are destroyed with slightly different rates.
}
\label{fig:73P_hcn}
\end{figure}

As a starting point for the validation of our time-dependent model described in Sect.~\ref{sect:rad}, we adopt the case for comet 73P/Schwassmann-Wachmann, hereafter 73P, which broke into fragments during its 2006 apparition. \citet{Paganini2010} observed the HCN $J=3-2$ and
$J=4-3$ lines in 73P on May 12 2006 using the Heinrich Hertz Submillimeter Telescope (HHSMT). Contemporary Caltech Submillimeter Observatory (CSO) HCN observations of 73P/B were also made \citep{Lis2008}.
\citet{Paganini2010} also performed comprehensive modelling of the HCN emission to which we can compare our modelling results. We adopt the values for fragment B these authors used: $Q_\mathrm{HCN} = 3\times 10^{25}\,\mathrm{mol\, s^{-1}}$,
$Q_\mathrm{H_2O} = 1.9\times 10^{28}\,\mathrm{mol\, s^{-1}}$, $T=90\,\mathrm{K}$, and $v_\mathrm{e}=0.53\,\mathrm{km\, s^{-1}}$. To further mimic their setup we neglect hfs and include effective
pumping rates. In the upper panel of Fig.~\ref{fig:73P_hcn} we show the HCN fractional level populations for $J=2, 3, 4$ as a function
of cometocentric radius.

The appearance of the level population variation looks very similar to that of
Fig.~1 in \citet{Paganini2010} hinting at any time-dependent effects not being huge in this particular case. Our late time values are about a factor 1.2-1.5 larger
than those of \citet{Paganini2010}. The middle panel in this figure depicts the relative importance of the different
HCN excitation mechanisms (for $J=3-2$). In the inner coma, collisions with water molecules are most important while
in the outer coma, radiative processes dominate. It between, as shown by \citet{Xie1992},
collisions by electrons are the most important excitation mechanism.

The different downward rates in the middle panel of Fig.~\ref{fig:73P_hcn} have been normalized
with the rate due to expansion and destruction, $\gamma$ in Eq.~(\ref{eq:gamma}). When this ratio is around 1 or smaller,
we may see time dependency effects. As visualized in the bottom panel of Fig.~\ref{fig:73P_hcn}, where the radial density
variations of water, HCN and electrons are shown, the electron properties change very rapidly at the
contact surface radius, $R_\mathrm{CS} = 324\,\mathrm{km}$. Here the electron temperature $T_\mathrm{e}$ rises very steeply
from the comet temperature to around $10^4$ K \citep{Ip1985, Bensch2004} accompanied with an increase of the electron
density.

\begin{figure}
\centering
\includegraphics[width=8.3cm]{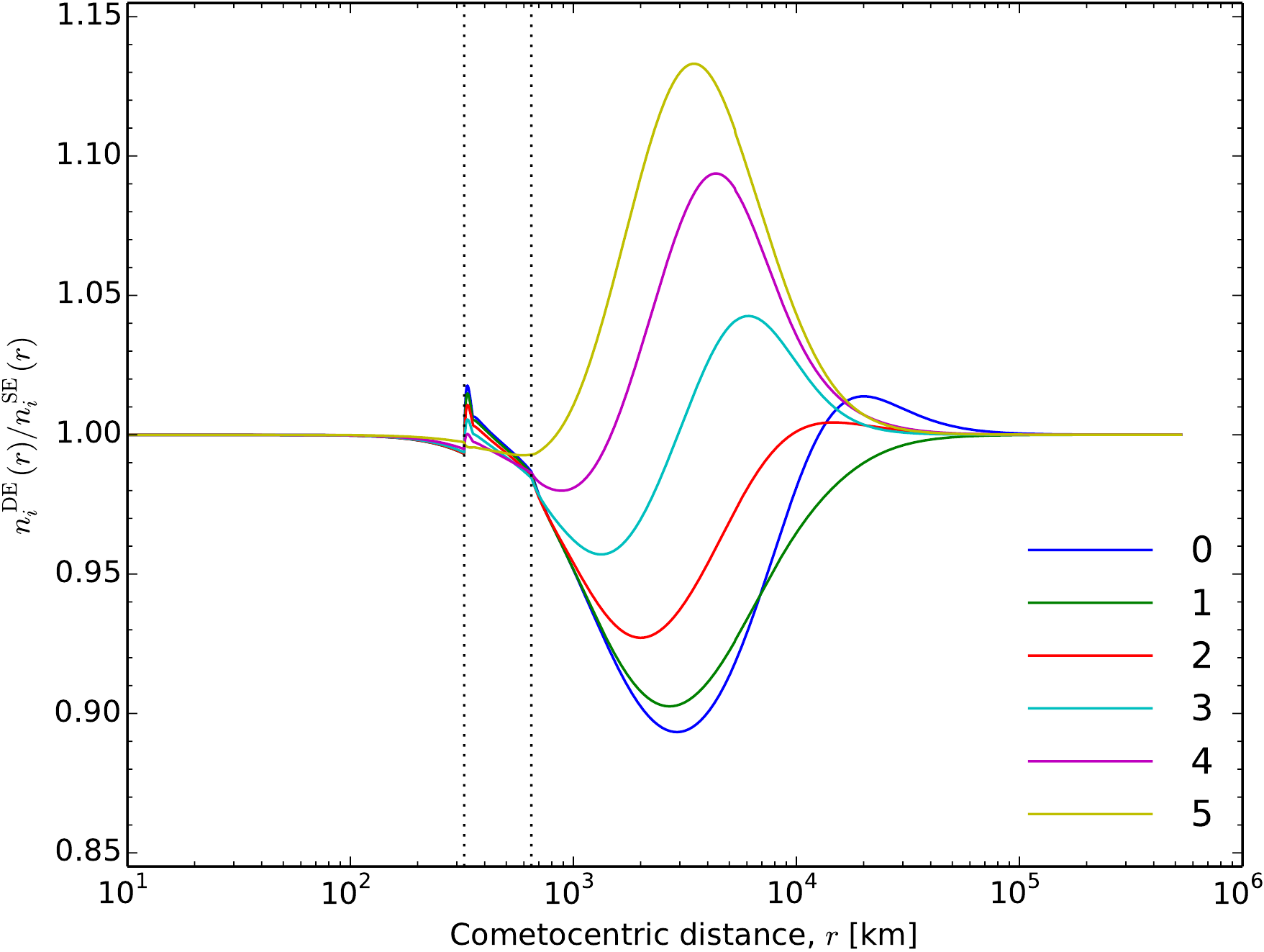}
\caption{The HCN level ($J=0$ to 5) population variation as a function of cometocentric radius, for the time-dependent
modelling (see Fig.~\ref{fig:73P_hcn}) adopting the parameters for fragment B of 73P/Schwassmann-Wachmann during its 2006
passage. The level populations have been normalized to their SE-values to visualize
time-dependent effects. The vertical dotted lines correspond to radii $R_\mathrm{CS}$ and $2R_\mathrm{CS}$ between which
the electron temperature increases from 90~K to $10^4$~K.
}
\label{fig:73P_rpop}
\end{figure}

It is around a cometocentric radius near $R_\mathrm{CS}$, see the middle panel in Fig.~\ref{fig:73P_hcn}, where we should start to see the
time-dependent effects on the excitation \citep[cf.][]{Chin1984}. To investigate this in more detail, we display in Fig.~\ref{fig:73P_rpop} the ratio of
the time dependent level populations ($J=0$ to 5) to those valid in the SE-limit. The level
population ratios are very close to 1 up to about $r=100\,\mathrm{km}$ \citep[cf.][]{BM1984} and from radii larger than about
$r=(4-5)\times 10^4\,\mathrm{km}$. After the sharp increase in $T_\mathrm{e}$ from 90~K to $10^4$~K over the interval
$[R_\mathrm{CS}, 2R_\mathrm{CS}]$ the level populations deviate from the SE-values during the settling -- in this case at most up
to 7\% for $J=2$. The low-$J$ level populations settle a little later than do the high-$J$ level populations since their $A$-coefficients
are smaller. This deviation is not directly related to the change in electron properties but to that of the normalized rates, middle panel in Fig.~\ref{fig:73P_hcn}, which are close to 1. Due to its large electric dipole moment, HCN rotational transitions have large $A$-coefficients and we expect that
for molecules with smaller spontaneous rates (eg. CO) the settling of the level populations will take place further
out in the coma. The deviations from the SE populations will be larger if $Q_\mathrm{H_2O}$ is reduced since HCN-H$_2$O collisions
will be less effective in settling the populations. For instance, using $Q_\mathrm{H_2O} = 1\times 10^{28}\,\mathrm{mol\, s^{-1}}$ in our test case, the time-dependent $J=2$ population deviations reach 9\% and for $Q_\mathrm{H_2O} = 4\times 10^{28}\,\mathrm{mol\, s^{-1}}$ they are at the 5\% level. By reducing the amount of electron collisions (by lowering $x_{n_\mathrm{e}}$
from 0.3 to 0.15) the effects are less clear. Firstly, the initial time-dependent deviation at $R_\mathrm{CS}$ will be less pronounced
but the settling at larger radii, $10^3-10^4$~km, will show larger deviations. Increasing the
amount of electron collisions will reduce the slower time-dependent behaviour (and, of course, increase the
deviation at $R_\mathrm{CS}$).

\end{appendix}

\end{document}